\documentclass[12pt,recno]{article}

\usepackage{graphicx,amsfonts,amssymb,stmaryrd,amsmath,minRG,longtable}
\advance\hoffset by -1cm
\title{\vspace{-3cm}\begin{flushright}{\small LMU -- ASC 36/09  \\ \vspace{-0.4cm}RUNHETC-2009-17}\end{flushright}\vspace{2cm}\large \bf Defect Perturbations in Landau-Ginzburg Models
\vspace*{0.5cm}}

\author{\normalsize
Ilka Brunner$^{1,2}$\thanks{\tt E-mail: Ilka.Brunner@physik.uni-muenchen.de},
Daniel Roggenkamp$^3$\thanks{\tt E-mail: roggenka@physics.rutgers.edu},
Sebastiano Rossi$^{4}$\thanks{\tt E-mail: rossise@phys.ethz.ch}, 
\\ \\
${}^{1}${\small Arnold Sommerfeld Center, 
Ludwig Maximilians Universit\"at} \vspace*{-0.1cm} \\
{\small Theresienstr. 37, 
80333 M\"unchen, Germany} \vspace{0.3cm} \\
${}^{2}${\small Excellence Cluster Universe, 
Technische Universit\"at M\"unchen} \vspace*{-0.1cm} \\
{\small Boltzmannstr. 2, 
85748 Garching, Germany} \vspace{0.3cm} \\
${}^3$ {\small Department of Physics and Astronomy, Rutgers University}
\vspace{-0.1cm} \\
{\small Piscataway, NJ 08855-0849} \vspace{0.3cm}\\
${}^{4}${\small Institut f\"ur Theoretische Physik, 
ETH Z\"urich} \vspace*{-0.1cm} \\
{\small 8093 Z\"urich, Switzerland} \vspace{0.3cm} \\
}

\begin{document}
\maketitle

\begin{abstract}
Perturbations of B-type defects in Landau-Ginzburg models are considered. 
In particular, the effect
of perturbations of defects on their fusion is analyzed in the framework of matrix factorizations. As an application, it is discussed how fusion with perturbed defects induces perturbations on boundary conditions. It is shown that in some classes of models all boundary perturbations can be obtained in this way. Moreover, a universal class of perturbed defects is constructed, whose fusion under certain conditions obey braid relations. The functors obtained by fusing these defects with boundary conditions
are twist functors as introduced in the work of Seidel and Thomas.
\end{abstract}

\thispagestyle{empty}

\newpage

\tableofcontents
\newpage


\section{Introduction}

A defect in a two-dimensional field theory is a line of inhomogeneity on the
surface on which the theory is defined. In general defect lines carry extra degrees of freedom not inherited from the bulk, which determine how excitations are transmitted between the theories 
on either side\footnote{The theories on the two sides of the defect
can either be the same or different. In case the theories are different, defects are sometimes referred to as ``interfaces''.}. 

The theory of defects is closely related to the theory of boundary conditions. Consider for example two conformal field theories on the complex plane separated by a defect located along the real line. Folding the plane along this line results in a surface with boundary for a ``doubled theory'', which is given by
the theory on the upper half plane tensored by the conjugate of the theory on the lower half plane \cite{WOng:1994pa,Bachas:2001vj}. 

However, compared to boundary conditions, defects have more structure: They can form junctions, and they can be composed by {\it fusion}. Fusion is the process in which two parallel defects are brought infinitely close together
\cite{Petkova:2000ip,Frohlich:2006ch,Runkel:2007wd,Bachas:2007td}. 
In the limit, a new defect is created. 
Of course, taking such a limit is in general a highly singular procedure, which in special situations simplifies however.
Most notably, there are so-called topological defects that preserve the full diffeomorphism invariance. They can be moved around freely, in particular without causing any singularities. 
Hence, they can be fused smoothly. 

In the context of supersymmetric $N=(2,2)$ models one
can consider defects preserving A- or B-type supersymmetry. These defects survive the corresponding
topological A- or B-twists respectively. On the level of the twisted theory,
their fusion is regular and defines
a product structure on all such defects \cite{Brunner:2007qu}. 

Defects and their fusion have a variety of applications in the context of string
theory and conformal field theory. In the string theory context, it was proposed in \cite{Bachas:2008jd} that defects can be used as spectrum generating symmetries. The main idea is that fusion of a conformal boundary condition describing a D-brane in CFT1
with a topological defect between CFT1 and CFT2, produces a conformal boundary condition in CFT2. Since conformal invariance is equivalent to the classical string equations, fusion with topological defects
creates new solutions of classical string theory out of given ones.

Certain special defects arise between UV and IR fixed points of  quantum field theories \cite{Brunner:2007ur}. These defects can be used to describe how boundary conditions behave under the corresponding renormalization group flows. In this way, defects and their fusion can serve as 
an alternative to the perturbative analysis of this problem.
This has been made explicit in the case of $N=(2,2)$ minimal models in \cite{Brunner:2007ur}\footnote{Recently, the
paper \cite{Fredenhagen:2009tn} explored the possibility to use certain topological defects to investigate bulk-boundary flows on the level of the full conformal field theory.}. 

In the same spirit, defects can be used to describe D-brane monodromies \cite{Brunner:2008fa}. Namely, there are 
defects associated to exactly marginal bulk deformations as well. 
Fusion with a defect associated to a deformation along a closed loop in the bulk moduli space encodes the effect of the corresponding monodromy on boundary conditions. 

But defects are not only useful in the analysis of bulk perturbations of theories with boundary. They also
relate different boundary renormalization group flows \cite{Graham:2003nc,Bachas:2004sy,Alekseev:2007in}. The basic idea is that certain boundary perturbations can be pulled back to the bulk by splitting off defects. In the context of WZW models this has been analyzed in  \cite{Bachas:2004sy}.  Turning things around, fusing a perturbed defect with a boundary condition, the defect perturbation descends to a boundary perturbation of the boundary condition emanating from the fusion.
Hence, defect perturbations give rise to classes of perturbations of different boundary conditions.

Motivated by these observations, in this paper, we will study perturbed defects and their fusion in $N=(2,2)$ supersymmetric theories, in particular Landau-Ginzburg models
(see \cite{Manolopoulos:2009np,Kormos:2009sk,Fredenhagen:2009tn} for recent related work in conformal field theory). In Landau-Ginzburg models, B-type defects have a convenient realization in terms of matrix factorizations  \cite{Brunner:2007qu}. Their fusion is regular and essentially given by
a tensor product between Chan-Paton type spaces. Therefore, this framework lends itself easily to the analysis of fusion of perturbed defects. This will be used to discuss how defect perturbations induce boundary perturbations in the way alluded to above. 

In the case of the Landau-Ginzburg models with one chiral superfield and superpotential $W=x^d$ we establish that all supersymmetry preserving boundary perturbations arise in this way, \ie all boundary perturbations can be pulled back into the bulk by means of defects. The same applies to $\ZZ_d$-orbifolds of these models, and to
models which are tensor products of two identical models. Although we treat it in the Landau-Ginzburg framework, we expect that the arguments in the latter case generalize to tensor products of arbitrary $N=(2,2)$ theories with their conjugates. 

In the IR, the Landau-Ginzburg orbifolds with superpotential $W=x^d$ are described by 
$N=2$-supersymmetric minimal models, which are rational conformal field theories with diagonal\footnote{with respect to B-type supersymmetry} modular invariants. Thus, in these models we can relate our considerations to results obtained for diagonal RCFTs.
Of course, 
defect perturbations are much more difficult to deal with on the level of the full conformal field theory, but some special classes of perturbations have been treated in \cite{Runkel:2007wd}. 

Apart from the induction of boundary perturbations, we use defect perturbations
to construct 
special classes of defects which show 
an interesting universal behavior. 
More precisely, in any theory there are purely reflective defects which impose fixed boundary conditions on the two theories on either side, as well as the ``trivial" or identity defect. 
Between these defects there is a canonical defect changing field, which descends from the identity
field on the boundary condition imposed by the reflective defect. It can be used
to perturb superpositions of reflective and identity defects.
We show that the resulting perturbed defects have the following nice properties.

If the underlying boundary condition is ``spherical'' in the sense that the BRST-cohomology of  boundary fields on it is two-dimensional and the boundary two-point functions are non-degenerate, then 
the associated defect is group-like as defined in \cite{Frohlich:2004ef}, \ie the defect and its dual fuse to the identity defect. Furthermore these defects obey a twisted commutation relation with respect to fusion. 
  
Moreover, if there is a collection $(P_1,\ldots,P_m)$ of spherical boundary conditions, such that there 
is exactly one BRST-invariant boundary condition changing field between any neighboring $P_i$, $P_j$, $|i-j|=1$ and no one between $P_i$ and $P_j$, $|i-j|>1$, \ie the spherical boundary conditions form an $A_m$-sequence\footnote{Examples include the $A_m$ chains on K3 surfaces  responsible for the non-abelian gauge symmetries of type II strings.},
the associated defects satisfy braid relations with respect to fusion.

We carry out the construction and discussion in the context of Landau-Ginzburg models, but we expect it to be valid in any $N=(2,2)$ supersymmetric theory.   

In fact, these defects are generalizations of defects describing mo\-no\-dro\-mies around conifold points \cite{Brunner:2008fa}. Their fusion with boundary conditions provides a world sheet realization of the twist functors introduced in the construction of braid group representations in the group of autoequivalences of certain categories in \cite{Seidel-Thomas}. 

This paper is organized as follows: In Section \ref{sec:MF} we review the matrix factorization formalism used to describe B-type defects and boundary conditions in Landau-Ginzburg models.
Section \ref{sec:perturb} contains a general discussion of perturbed defects and their fusion in this framework.
Section \ref{sec:boundaries} is devoted to the fusion of perturbed defects with boundary conditions, and the induced boundary perturbations. Finally in Section \ref{sec:braidgroup} we construct the universal twist defects and establish that under some conditions their fusion satisfies braid relations.
 We provide various classes of examples in which these conditions are satisfied.

\section{Brief review of matrix factorizations}\label{sec:MF}

In Landau-Ginzburg models,  B-type supersymmetric D-branes as well as B-type supersymmetry preserving defects have an elegant description
in terms of matrix factorizations \cite{Kapustin:2002bi,Brunner:2003dc,Khovanov:2004bc,Brunner:2007qu}, see \cite{Jockers:2007ng,Knapp:2007vc} for reviews.

 A matrix factorization $P$ of a polynomial
$W\in\CC[x_1,\ldots,x_N]$ is given by a pair $(P_1,P_0)$ of free $\CC[x_1,\ldots,x_N]$ modules together with homomorphisms $p_s:P_s\rightarrow P_{(s+1)\,{\rm mod}\, 2}$ between them 
which compose to $W$ times the identity map, \ie $p_1p_0=W\id_{P_0}$ and $p_0p_1=W\id_{P_1}$. In the following we will often represent matrix factorizations by
\beq
P: P_1\overset{p_1}{\underset{p_0}{\rightleftarrows}}P_0\,.
\eeq
Sometimes it is useful to regard them as two-periodic twisted\footnote{The differential squares to $W$ instead of zero.} complexes. Indeed, such matrix factorizations form a category, with morphisms $\HH^*(P,Q)$ between two matrix factorizations $P$ and $Q$ given by the cohomology of the Hom-complex of the two twisted complexes associated to $P$ and $Q$. The latter is a two-periodic untwisted complex.  
\beq
\HH^* (P,Q)= H^*(\Hom(P,Q))\cong H^*(P^*\otimes Q)
\eeq
Here, $P^*$ denotes the dual matrix factorization
\beq\label{Mdual}
P^*: P_1^*\overset{p_0^*}{\underset{-p_1^*}{\rightleftarrows}}P_0^*\,,
\eeq
and the tensor product is the ordinary tensor product of complexes. It will be spelled out explicitly
in \eq{eq:MFtensor} below.

There are always matrix factorizations with modules $P_s=\CC[x_1,\ldots,x_N]$ and with maps
$p_r=1$ and 
$p_{(r+1)\,{\rm mod}\,2}=W$. They are trivial in the sense that they only have zero-morphisms with any other (including themselves) matrix factorization. Two matrix factorizations which differ by the addition of such a trivial matrix factorization are equivalent. Indeed, if a matrix representing one of the maps $q_i$ of a matrix factorization $Q$ contains a scalar entry different from zero, such a trivial matrix factorization can always be split off from $Q$.  

More generally, two matrix factorizations $Q$ and $Q'$ are equivalent
if one can find maps $u_i: Q_i \to Q_i'$ and $v_i: Q_i' \to Q_i$ such
that
\beq\label{Mequiv}
q\p_1=u_0q_1v_1\,,\quad q\p_0=u_1q_0v_0\,,\quad
q_1=v_0q\p_1u_1\,,\quad q_0=v_1q\p_0u_0
\eeq
and
\beqn
&&v_0u_0=\id_{Q_0}+\chi_1q_0+q_1\chi_0\,,\quad
v_1u_1=\id_{Q_1}+q_0\chi_1+\chi_0q_1\,,\\
&&u_0v_0=\id_{Q\p_0}+\chi\p_1q\p_0+q\p_1\chi\p_0\,,\quad
u_1v_1=\id_{Q\p_1}+q\p_0\chi\p_1+\chi\p_0q\p_1\,,\nonumber
\eeqn
Physically, the defects constructed using matrix factorizations can be regarded
as composites of a defect-anti-defect pair with a tachyon turned on. The data of the matrix factorization can be summarized in a defect BRST-operator 
\beq\label{eq:BRSTformulation}
{\cal Q}= \left( \begin{array}{cc} 0 & p_1 \\ p_0 & 0 \end{array} \right) 
\eeq
containing the tachyon profile.
This is an operator in $\End(P_1\oplus P_0)$, which is odd with respect to the $\ZZ_2$-grading 
\beq
\sigma=\id_{P_0}-\id_{P_1}\,.
\eeq
In this language, $\HH(P,Q)$ is just given by the BRST-cohomology on $\Hom(P,Q)$, and the equivalence relation (\ref{Mequiv}) becomes
\beq
{\cal Q}'= U {\cal Q} V, \quad UV =\id' + \{{\cal Q}, O' \}, \quad VU= \id + \{\ {\cal Q}, O \}
\eeq
for some $O$ and $O'$.

As was shown in \cite{Kapustin:2002bi,Brunner:2003dc,Lazaroiu:2003zi}, B-type supersymmetric D-branes in Landau-Ginzburg models with chiral superfields $x_1,\ldots,x_N$ and 
superpotential $W\in\CC[x_1,\ldots,x_N]$ can be represented by matrix factorizations of $W$, where open strings between two such D-branes are described by morphisms
between the respective matrix factorizations. 

In the same way, it has been argued in \cite{Brunner:2007qu} that B-type supersymmetry preserving defects between two Landau-Ginzburg models, one with chiral fields
$x_1,\ldots,x_N$ and superpotential $W_1\in\CC[x_1,\ldots,x_N]$ and one with chiral superfields $y_1,\ldots,y_M$ and superpotential $W_2\in\CC[y_1,\ldots,y_M]$
can be represented by matrix factorizations of $W_1-W_2$ over the polynomial ring $\CC[x_1,\ldots,x_N,y_1,\ldots,y_N]$.

As mentioned before, one interesting property of $N=2$-supersymmetric defects is that they can be fused with other such defects or boundary conditions preserving the same supersymmetry. Namely, two such defects can be brought on top of each other to produce a new defect, or a defect can be moved onto a world sheet boundary to change the boundary condition imposed there.
This fusion has a very simple realization in terms of the matrix factorization description. For instance, let $x_i$, $y_i$, $z_i$ be the chiral superfields of three Landau-Ginzburg models with superpotentials $W_1\in\CC[x_i]$, $W_2\in\CC[y_i]$ and $W_3\in\CC[z_i]$ respectively, which are separated by two 
defects represented by matrix factorizations $P^1$ of $W_1-W_2$ and $P^2$ of $W_2-W_3$. Fusing the two defects gives rise to a new defect separating the Landau-Ginzburg model with chiral fields $x_i$ and superpotential $W_1$ from the one with chiral fields $z_i$ and superpotential $W_3$. This fused defect is given by the matrix factorization
\beq\label{eq:fusion}
P^1*P^2=\left(P^1\otimes P^2\right)_{\CC[x_i,z_i]}^{\rm red}\,.
\eeq
Here, the tensor product of two matrix factorizations is defined by taking the tensor product of the associated twisted complexes. It is also a two-periodic complex which is twisted by the sum of the twists of the tensor factors. More concretely, the tensor product $P\otimes Q$ of matrix factorizations $P$ and $Q$ of $W$ and $W\p$ respectively can be written as
\beq\label{eq:MFtensor}
P\otimes Q: P_1\otimes Q_0\oplus P_0\otimes Q_1\overset{r_1}{\underset{r_0}{\rightleftarrows}}P_0\otimes Q_0\oplus P_1\otimes Q_1
\eeq
with 
\beq
r_1=\left(\begin{array}{cc} p_1\otimes\id&\id\otimes q_1\\-\id\otimes q_0&p_0\otimes\id\end{array}\right)\,,\quad
r_0=\left(\begin{array}{cc} p_0\otimes\id&-\id\otimes q_1\\ \id\otimes q_0&p_1\otimes\id\end{array}\right)
\eeq
which is a matrix factorization of $W+W\p$.

In the situation above, $P^1$ is a matrix factorization of $W_1-W_2$ and $P^2$ one of $W_2-W_3$.
Hence, $P^1\otimes P^2$ is a matrix factorization of $W_1-W_3\in\CC[x_i,z_i]$, but it is still a matrix factorization over $\CC[x_i,y_i,z_i]$. That means that the modules $(P^1\otimes P^2)_s$ are free $\CC[x_i,y_i,z_i]$-modules and also the maps 
$r_s$ between them depend on the $y_i$. 
The notation $\left(P^1\otimes P^2\right)_{\CC[x_i,z_i]}$ means that this matrix factorization has to be regarded as one over $\CC[x_i,z_i]$ only\footnote{In the following, it will usually be evident which base ring is chosen for matrix factorizations. For ease of notation we will therefore omit subscripts like $\CC[x_i,z_i]$ in equation \eq{eq:fusion}.}.
 As such, it is of infinite rank, because the modules $(P^1\otimes P^2)_s$ regarded as modules over $\CC[x_i,z_i]$ 
are free modules of infinite rank. For instance, $\CC[x_i,y_i,z_i]$ can be decomposed as 
\beq
\CC[x_i,y_i,z_i]=\bigoplus_{(l_1,\ldots,l_N)\in\NN_0^N}y_1^{l_1}\ldots y_N^{l_N}\CC[x_i,z_i]
\eeq
into free $\CC[x_i,z_i]$-modules.
Physically speaking, the chiral fields $y_i$ of the theory squeezed in between the two defects are promoted to new defect degrees of freedom in the limit where the two defects coincide. However, most of them are trivial. Namely, if both $P^1$ and $P^2$ are of finite rank, 
the matrix factorization $\left(P^1\otimes P^2\right)_{\CC[x_i,z_i]}$ can be reduced to finite rank by splitting off infinitely many trivial matrix factorizations. It is the result of this reduction $\left(P^1\otimes P^2\right)_{\CC[x_i,z_i]}^{\rm red}$ which describes the fused defect.

In the same way, fusion of B-type defects and B-type boundary conditions in Landau-Ginzburg models can be formulated in the matrix factorization framework. The fusion of a B-type defect separating a Landau-Ginzburg model with chiral fields $x_i$ and superpotential $W_1\in\CC[x_i]$ from one with chiral fields $y_i$ and superpotential $W_2\in\CC[y_i]$ and a B-type boundary condition in the second of these Landau-Ginzburg models can be represented by the matrix factorization
\beq
P*Q=\left(P\otimes Q\right)_{\CC[x_i]}^{\rm red}\,,
\eeq
where $P$ is the matrix factorization of $W_1-W_2$ associated to the defect and $Q$ the matrix factorization of $W_2$ associated to the boundary condition.

Certain defects are quite universal and exist in any Landau-Ginzburg model, or even any two-dimensional QFT. A special
example is the identity defect $\Id$. It is trivial in the sense that inserting it does not change any correlation functions. Nevertheless, it will play an interesting role in this paper. In Landau-Ginzburg models, it is realized by the following matrix factorization. The difference of the same superpotential in different variables can always be factorized as
\beq\label{eq:idfact}
W(x_i)-W(y_i) = \sum_i (x_i-y_i) A_i(x_i, y_i)\,.
\eeq
Denoting the rank-one factorizations with factors $p_1^{(i)}=x_i-y_i$ and $p_0^{(i)}=A_i(x_j,y_j)$ by
$\Id^{(i)}$, a matrix factorization representing the identity defect in the Landau-Ginzburg model with superpotential $W$ is given by the tensor product
\beq
\Id= \bigotimes \Id^{(i)}\,.
\eeq

Even though the factorization \eq{eq:idfact} is not unique, the equivalence class of $\Id$ is unique. Different choices of the $A_i$ lead to equivalent matrix factorizations $\Id$. 
 Indeed, as expected from the identity defect, fusion with the matrix factorization $\Id$ is trivial: $\Id * Q=Q$ \cite{Kapustin:2004df,Brunner:2007qu}. 

Another universal class of defects are totally reflective defects. Such defects provide boundary conditions for the theories on either side, and do not allow any excitations to be transmitted between the theories. In the context of matrix factorizations, such defects are realized by tensor products of matrix factorizations
of the superpotentials on the two sides. 
Let $P$ be a matrix
factorization of $W(x_i)$ and $Q$ of $ W(y_i)$. A totally reflective defect
imposing boundary condition $Q$ on one side and $P$ on the other side is
given by
\beq
T_{P,Q^*} := P\otimes Q^*,
\eeq
where the dual matrix factorization $Q^*$ was defined in (\ref{Mdual}). It arises
here because of the different orientations on the two sides of the defect. 
The fusion of $T_{P,Q^*}$ with matrix factorizations $R$ of $W(y_i)$ have a simple form
\beq
T_{P,Q^*}*R \equiv P\otimes (Q^* \otimes R) \equiv P\otimes H^*(Q^*\otimes R)
\equiv P\otimes {\cal H}^*(Q,R)\,.
\eeq
Here, the matrix factorization $P\otimes(Q^*\otimes R)$ has to be regarded as a matrix factorization
over $\CC[x_i]$. The factor $Q^*\otimes R$ is a factorization of $W(y_i)-W(y_i)=0$. It is
a complex with a regular differential $\delta$, which squares to zero and does not depend on the $x_i$. 
Thus, the non-zero matrix entries of $\delta$ contribute scalars to the matrix factorization $P\otimes(Q^*\otimes R)$ which can be used to reduce it to $P\otimes H^*(Q^*\otimes R)$.  

In Section \ref{sec:braidgroup} we will perturb the superposition of the identity and totally reflective defects and show that the resulting defects give rise to some interesting structures.

Everything described above for Landau-Ginzburg models easily carries over to Landau-Ginzburg orbifolds \cite{Brunner:2007ur}. If $\Gamma$ is a finite group acting on the ring $\CC[x_1,\ldots,x_N]$ of chiral fields of a Landau-Ginzburg model in such a way that the superpotential $W$ is $\Gamma$-invariant, one can consider its $\Gamma$-orbifold. B-type defects and boundary conditions in such orbifold theories can be represented by $\Gamma$-equivariant matrix factorizations \cite{Ashok:2004zb,Hori:2004ja,Brunner:2007ur}. These are matrix factorizations $P$ together with representations $\rho_i$ of $\Gamma$ on $P_i$ which are compatible with the ring structure, and with respect to which the maps $p_i$ are invariant:
$\rho_{(i-1)\,{\rm mod}\,2}p_i=p_i\rho_i$. These conditions ensure that there is an induced representation of $\Gamma$ on the BRST-cohomology $\HH^*(P,Q)$ which can be used to define the
BRST-cohomology of the orbifold theory as the $\Gamma$-invariant part
\beq
\HH^*_{\rm orb}(P,Q)=\left(\HH^*(P,Q)\right)^{\Gamma}
\eeq
of the respective BRST-cohomology in the underlying unorbifolded model. 

Similarly, fusion of two defect matrix factorizations in the orbifold theory 
is given by the $\Gamma_{\rm squeezed}$-invariant
part of the fusion of the underlying matrix factorizations
\beq
P^1*_{\rm orb} P^2=\left(P^1*P^2\right)^{\Gamma_{\rm squeezed}}\,.
\eeq
Here, $\Gamma_{\rm squeezed}$ denotes the orbifold group of the Landau-Ginzburg model squeezed in between the two defects. For more details on B-type defects in Landau-Ginzburg orbifolds see \cite{Brunner:2007ur}.

\section{Perturbed defects and their fusion}\label{sec:perturb}

\subsection{Perturbation of defects}

The perturbation theory of defects exactly parallels the one of boundary conditions.
Interesting new effects arise however, when one considers structures inherent to defects, which are not present in boundary conditions, for instance fusion.
Let us start by briefly discussing perturbations of B-type defects in Landau-Ginzburg models, following \cite{Hori:2004ja}. 

Just like for boundaries, there are  fields which are confined to defects, and which can be used to perturb
the latter. In the context of Landau-Ginzburg models, supersymmetry preserving perturbations of B-type defects $P$ are generated by fields in $\HH(P,P)$, and they correspond to deformations of the corresponding matrix factorizations. In the BRST-formulation \eq{eq:BRSTformulation} such a deformation is a family
\beq\label{defectfamily}
{\cal Q}(t)={\cal Q}^0+\sum_{n>1} t^n\psi_n
\eeq
of BRST-operators with 
\beq\label{eq:BRSTdefcond}
({\cal Q}(t))^2=W
\eeq
 for all $t$. To first order in $t$ this condition means that $\psi_1$ has to be BRST-closed with respect to the undeformed BRST-operator ${\cal Q}^0$
\beq
\{{\cal Q}^0,\psi_1\}=0\,.
\eeq
On the other hand, if $\psi_1$ is BRST-closed with respect to ${\cal Q}^0$
\beq
\psi_1=\left[{\cal Q}^0,\chi_1\right]
\eeq 
then the first order deformation can be compensated by the equivalence
\beq
{\cal Q}\mapsto e^{-t\chi_1}{\cal Q}e^{t\chi_1}\,.
\eeq
Thus, to first order, deformations of (equivalence classes of) matrix factorizations $P$ are generated by 
$\HH^1(P,P)$. Of course, not all first order deformations are necessarily integrable.
In general, there can be obstructions at higher order, 
as has been analyzed in \cite{Ashok:2004xq,Hori:2004ja,Knapp:2006rd,Carqueville:2009ay}.  For instance at second order, condition \eq{eq:BRSTdefcond} implies
\beq\label{eq:BRSTdef2}
\psi_1^2+\{{\cal Q}^0,\psi_2\}=0\,.
\eeq
If $\psi_1$ squares to a non-trivial BRST-cohomology class, equation \eq{eq:BRSTdef2} cannot be satisfied and the deformation generated by $\psi_1$ is obstructed. Otherwise, one can find $\psi_2$ such that \eq{eq:BRSTdef2} holds. This can be repeated order by order: given $\psi_1,\ldots,\psi_{n-1}$ such that \eq{eq:BRSTdefcond} holds to order $n-1$ one has to construct $\psi_n$ such that it is satisfied to order $n$. If for some $n$ this is not possible, the deformation is obstructed. Otherwise
one obtains a family of non-equivalent BRST-operators parametrized by $t$, or to put it differently a family of non-equivalent matrix factorizations (see \cite{Hori:2004ja} for a more detailed discussion in case of boundaries).

A special case arises, when the undeformed matrix factorization is a direct sum of two matrix factorizations $P$ and $P'$, and the deformation is generated by a ``defect changing operator"
$T\in\HH^1(P,P')$. 
It describes the 
bound state formation of the associated defects.

Since in this case 
$T^2=0$, condition \eq{eq:BRSTdefcond} is automatically satisfied to all orders, and no higher order terms are necessary. In particular, all such deformations are 
integrable, and the family of matrix factorizations are given by the mapping cones
\beq
Q(t)=\cone(tT:P\longrightarrow P'):\; P_1\oplus P'_1\maplr{c_1}{c_0}P_0\oplus P'_0
\eeq
with
\beq
c_1=\left(\begin{array}{cc} p_1 & 0 \\ tT|_{P_1} & p'_1\end{array}\right)\,,\quad
c_0=\left(\begin{array}{cc} p_0 & 0 \\ tT|_{P_0} & p'_0\end{array}\right)\,.\nonumber
\eeq
Note that all $Q(t)$ for $t\neq 0$ are equivalent.

\subsection{Fusion of perturbed defects}

Our focus in this paper is the fusion of perturbed defects, in particular of those defects which can be obtained as bound states. Of course, the fusion product of a perturbed defect $D(t)$ with another defect $D'$ can be viewed as a perturbation of the fusion product of the unperturbed defect with the other defect
\beq
D(t)*D'=(D*D')(t)\,.
\eeq
Once the obstruction problem is solved for the initial defect $D$ and a given direction $\psi_1\in\HH^1(D,D)$, it is automatically solved for $D*D'$ for an induced direction $\widetilde{\psi}_1\in\HH^1(D*D',D*D')$. Indeed, this is obvious, because
fusing a family of defects with another defect one obtains again a family of defects.

The first question that arises
is how to determine the induced direction $\widetilde{\psi}_1$.
The answer is indeed straight-forward
to work out in the Landau-Ginzburg framework.

Let us start with an unperturbed defect corresponding to a matrix factorization
of $W_1(x_i)-W_2(y_i)$ with BRST-operator ${\cal Q}^0_1$.  Adding a perturbation generated by $\psi_1$ results in a perturbed BRST-operator ${\cal Q}_1(t)={\cal Q}^0_1+\sum_{n>0}t^n\psi_n$. 
We now take the fusion product with an arbitrary other defect with BRST-operator ${\cal Q}_2$, $({\cal Q}_2)^2=W_2(y_i)-W_3(z_i)$. Fusion creates a new defect with BRST-operator
\beq
{\cal Q}(t)=  {\cal Q}^0_1 + \sum_{n>0}t^n \psi_n + {\cal Q}_2= {\cal Q}^0 +\sum_{n>0}t^n\psi_n
\eeq
which correctly squares to $W_1(x_i)-W_3(y_i)$. 
This equation just reflects the fact that the fusion of a perturbed defect with another defect can be interpreted as a perturbation of the fusion of the unperturbed defects.
The BRST-operators ${\cal Q}(t)$ appearing above still depend on the chiral fields 
$y_i$, which in the fusion process were promoted to new defect degrees of freedom.
In terms of matrix
factorizations, the result is an infinite dimensional matrix factorization over $\CC[x_i,z_i]$. 
They can be made finite dimensional by equivalence transformations involving stripping off infinitely many trivial matrix factorizations. For the unperturbed fusion there exist polynomial  matrices $U,V$ that are inverse to each other up to BRST-equivalence and satisfy 
\beq
U {\cal Q}^0 V = [{\cal Q}^0]^{\rm red} \ ,
\eeq
where $[{\cal Q}^0]^{\rm red}$ is the finite dimensional reduction of ${\cal Q}^0$.
Once the equivalences $U,V$ are
determined for the unperturbed fusion,
they can be used to map the perturbing field $\psi_1$ and the higher order terms $\psi_n$, $n>1$ to the induced fields $\widetilde{\psi}_n$ on
$[{\cal Q}^0]^{\rm red}$.

As discussed above, in case the perturbed defect $D(t)$ corresponds to a bound state, the same is true for $D(t)*D'$. Hence, also the deformation problem for $D(t)*D'$ is solved at first order and all $\widetilde{\psi}_n=0$ for $n>1$. Only $\widetilde{\psi}_1$ needs to be determined.

Consider for instance the bound state
\beq
P=\cone(T:P^{(1)}\to P^{(2)})
\eeq
for some $T\in\HH^1(P^{(1)},P^{(2)})$.

The fusion with a defect represented by a matrix factorization
$Q$ then takes the form
\beqa
P * Q &=& \cone(T: P^{(1)} \to P^{(2)}) * Q \\ \nonumber
&=& \big[  \cone(T: P^{(1)} \to P^{(2)}) \otimes  Q \big]^{\rm red} \\ \nonumber
&=& \big[  \cone(T\otimes \id: P^{(1)}\otimes Q \to P^{(2)} \otimes  Q) \big]^{\rm red} \\ \nonumber
&=& \cone(\widetilde{T}:  P^{(1)}* Q \to P^{(2)} *  Q) 
\eeqa
In the last step, the tensor product matrices have been reduced to finite dimension,  
and the tachyon $T$ was transferred accordingly by means of the
equivalences $P^{(i)}\otimes Q\cong P^{(i)}*Q$,.

\subsection{Calculation of perturbed fusion products}\label{fusiontools}

As alluded to in Section \ref{sec:MF} the most difficult part in determining the fusion $P*Q=(P\otimes Q)^{\rm red}$ of two matrix factorizations $P$ of $W_1(x_i)-W_2(y_i)$ and $Q$ of $W_2(y_i)-W_3(z_i)$ is to reduce their tensor product to finite dimension. Indeed, trying to find the corresponding equivalences directly on the level of matrix factorizations can be very intricate.
However as put forward in \cite{Brunner:2007qu} one can make use of the relation between maximal Cohen-Macaulay modules and matrix factorizations \cite{Eisenbud}. Namely, instead of considering the matrix factorization $Q'=P\otimes Q$ over $\CC[x_i,z_i]$, one can equivalently consider an  $R=\CC[x_i,z_i]/(W_1(x_i)-W_3(z_i))$-module $V$ with 
a projective resolution 
\beq
\ldots \stackrel{v_{n+1}}{\longrightarrow} V_n
\stackrel{v_{n}}{\longrightarrow} V_{n-1}\stackrel{v_{n-1}}{\longrightarrow}\ldots\stackrel{v_1}{\longrightarrow} V_0=V\rightarrow 0\,,
\eeq
which after a finite number of steps turns into the two-periodic complex determined by the matrix factorization $Q'$
\beq
\begin{array}{ll}
V_{N+2i}= Q'_0/(W_1-W_3)Q'_0\,,&
V_{N+2i+1}= Q'_1/(W_1-W_3)Q'_1\,,\\
 v_{N+2i}=q'_0\,,&
v_{N+2i+1}=q'_1\,,\end{array}
\eeq
for all $i\geq 0$.
Instead of reducing $Q'$ one can now reduce $V$ to a finite rank $R$-module $\widetilde{V}\cong V$ and calculate a projective resolution
\beq
\ldots \stackrel{\widetilde{v}_{n+1}}{\longrightarrow} \widetilde{V}_n
\stackrel{\widetilde {v}_{n}}{\longrightarrow} \widetilde{V}_{n-1}\stackrel{\widetilde{v}_{n-1}}{\longrightarrow}\ldots\stackrel{\widetilde{v}_1}{\longrightarrow} \widetilde{V}_0=\widetilde{V}\rightarrow 0\,,
\eeq
which also turns two-periodic after a finite number of steps
\beq
\begin{array}{ll}
\widetilde{V}_{\widetilde{N}+2i}= S_0/(W_1-W_3)S_0\,,&
\widetilde{V}_{\widetilde{N}+2i+1}= S_1/(W_1-W_3)S_1\,,\\
\widetilde{v}_{\widetilde{N}+2i}=s_0\,,&
\widetilde{v}_{\widetilde{N}+2i+1}=s_1\,,\end{array}
\eeq
for all $i\geq 0$. Here the $S_i$ are free $\CC[x_i,z_i]$-modules of finite rank, and the two-periodic part of the resolution gives rise to a finite dimensional matrix factorization
\beq
S: S_1 \maplr{s_1}{s_2} S_0
\eeq
of $W_1(x_i)-W_3(z_i)$ over $\CC[x_i,z_i]$. 
The isomorphisms $r:V\rightarrow \widetilde V$ and $r^*:\widetilde{V}\rightarrow V$ lift to 
 the resolutions 
\beq
\begin{array}{cccccccc}
\ldots & \stackrel{v_3}{\longrightarrow} & V_2 & \stackrel{v_2}{\longrightarrow} &
V_1 & \stackrel{v_1}{\longrightarrow} & V_0\cong V & \rightarrow 0\\
  &&  \mapupdown{r_2}{r^*_2} &  &
 \mapupdown{r_1}{r^*_1} &  &  \mapupdown{r}{r^*} & \\
\ldots & \stackrel{\widetilde{v}_3}{\longrightarrow} & \widetilde{V}_2 & \stackrel{\widetilde{v}_2}{\longrightarrow} &
\widetilde{V}_1 & \stackrel{\widetilde{v}_1}{\longrightarrow} & \widetilde{V}_0\cong \widetilde{V} & \rightarrow 0\end{array}\,,
\eeq
and the $r_i$ and $r_i^*$ for $i>N,\widetilde{N}$ provide an equivalence of the matrix factorizations 
$Q'$ and $S$. In this way, one can obtain a finite dimensional matrix factorization $S$ equivalent to $Q'$, and also determine the equivalence between the two. The latter can in particular be used to map morphisms of $Q'$ to those of $S$. 

\section{An Application: Boundary flows from defects}\label{sec:boundaries}

A special case of the fusion processes discussed above is the fusion of a perturbed defect
with a boundary condition.
If the perturbation of the defect is unobstructed, so that it gives rise to a family of supersymmetry preserving defects, fusion with a boundary condition immediately yields a family of 
boundary conditions. The deformation problem on the boundary does not need to be solved again,
it is solved on the level of the defect.
Note that one and the same family
of defects can be fused with many different boundary conditions. This
means that unobstructed directions in the moduli space of different D-branes are in fact related: They are universal flat directions in the
sense that they can be pulled back to the bulk using the same defect.

This holds in particular for defects, which can be obtained as bound states. Fusing such a defect with 
a boundary condition yields a bound state of boundary conditions, where the tachyon is induced by the 
one on the defect. This again means that certain tachyon condensation
processes of D-branes are universal in the above sense and can be pulled back to the bulk, as has been discussed for WZW models in \cite{Bachas:2004sy}. 
The advantage of the Landau-Ginzburg language is that
the fusion product can easily be calculated, and that it is therefore straight forward to 
determine the resulting boundary flows. We will illustrate this in some
examples.

\subsection{Example: Minimal models}\label{sec:mmbranes}

\subsubsection{B-type boundary conditions in minimal models}

Consider a Landau-Ginzburg model with superpotential 
\beq
W=x^d \ .
\eeq
B-type supersymmetric boundary conditions in these models can be represented by matrix factorizations
of $W$. All matrix factorizations of $W$ can be obtained as cones of the elementary matrix factorizations
\beq
Q^\ell:Q^\ell_1=\CC[x]\maplr{x^\ell}{x^{d-\ell}}\CC[x]=Q^\ell_0\,.
\eeq
As described in Section \ref{sec:MF}, the
 open string spectrum between two different boundary conditions can be obtained 
  as the BRST-cohomology $\HH(Q^{\ell_1},Q^{\ell_2})$ of the respective matrix factorizations $Q^{\ell_i}$
\begin{center}
\begin{picture}(400,110)(-100,20)
\put(00,100){$Q^{\ell_1}:$}
\put(40,100){$\CC[x]$}
\put(80,108){\vector(1,0){50}}
\put(105,113){\footnotesize{$x^{\ell_1}$}}
\put(150,100){$\CC[x]$}
\put(130,100){\vector(-1,0){50}}
\put(105,90){\footnotesize{$x^{d-\ell_1}$}}
\put(00,30){$Q^{\ell_2}:$}
\put(40,30){$\CC[x]$}
\put(80,38){\vector(1,0){50}}
\put(105,43){\footnotesize{$x^{\ell_2}$}}
\put(150,30){$\CC[x]$}
\put(130,30){\vector(-1,0){50}}
\put(105,20){\footnotesize{$x^{d-\ell_2}$}}
\put(50,90){\vector(0,-1){45}}
\put(160,90){\vector(0,-1){45}}
\put(30,65){$\phi_1$}
\put(170,65){$\phi_0$}
\put(70,90){\vector(2,-1){75}}
\put(140,90){\vector(-2,-1){75}}
\put(75,75){$t_1$}
\put(125,75){$t_0$}
\end{picture}
\end{center} 
A BRST-invariant fermion $t=(t_1,t_0):Q^{\ell_1}\to Q^{\ell_2}$ has to satisfy 
\beq
x^{d-\ell_2}t_1 + t_0 x^{\ell_1} =0\,.\nonumber
\eeq
In the case that $\ell_1 + \ell_2 \leq d$ we can solve for $t_0$
\beq
t_0= - t_1 x^{d-\ell_1-\ell_2} \ ,
\eeq
otherwise, if $\ell_1+\ell_2 >d$
\beq
t_1=-t_0 x^{\ell_1+\ell_2 -d}
\eeq
The fermion is BRST-exact if
\beq
t_1 = x^{\ell_2} \phi_1 + \phi_0 x^{\ell_1}, \quad
t_0 = \phi_0 x^{d-\ell_2} + \phi_1 x^{d-\ell_1} \,.
\eeq
Hence, the fermionic BRST-cohomology can be described as
\beq\label{eq:fermboundary}
t=(t_1,t_0)=(t_1,-t_1x^{d-\ell_1-\ell_2})\,,\;\;t_1\in x^b\CC[x]/\langle x^a\rangle\cong\HH^1(Q^{\ell_1},Q^{\ell_2})\,,
\eeq
where 
$a={\rm min}\{\ell_1,\ell_2\}-1$ and $b={\rm max}\{d-\ell_1-\ell_2,0\}$.

The possible tachyon condensation processes have been described in the Landau-Ginzburg framework in \cite{Hori:2004bx}. Deformations of a single $Q^\ell$ are not integrable, but perturbations with defect changing fields are.  As discussed above, they can be represented by cones
\beq
 \cone (t: Q^{\ell_1} \to Q^{\ell_2} )
\eeq
In the following we will demonstrate that all these perturbations are induced by fusion with perturbed defects. 
The idea is to generate all boundary conditions by fusing defects with the boundary condition
corresponding to the elementary matrix factorization $Q^1$, and to show that the boundary spectra 
can be induced from the defects. Let us start by introducing the defects which we will use. 

\subsubsection{B-type defects in minimal models}\label{sec:mmdefects}

B-type defects in minimal models can be represented by matrix factorizations of the superpotential
\beq
W=x^d-y^d\,.
\eeq
A nice class of such defects can be easily obtained by grouping the linear factors of
\beq
W(x)-W(y)=\prod_{l=1}^d (x-\eta_l y), \quad \eta_l=e^{\frac{2\pi i l}{d}}
\eeq
into two sets. One obtains rank-one factorizations
\beq\label{idartig}
P^{I}: \CC[x,y] \maplr{p_1^I}{p_0^I}  \CC[x,y]
\eeq
with
\beq
p_1^I = \prod_{i\in I} (x-\eta_i y), \quad p_0^I = \prod_{i\in D\setminus I} (x-\eta_i y) ,
\eeq
and  $D=\{1, \dots, d\}$ and $I \subset D$. 

\subsubsection{Inducing boundary flows by defects}

It is not difficult to see \cite{Brunner:2007qu} that an elementary matrix factorization $Q^\ell$ can be obtained by fusing any defect matrix factorization $P^I$ with $|I|=\ell$ with $Q^1$
\beq
P^I*Q^1\cong Q^{|I|}\,.
\eeq
Namely, as described in Section \ref{fusiontools} to reduce the infinite dimensional matrix factorization 
$Q'=P^I\otimes Q^1$ we consider the $R=\CC[x]/\langle x^d\rangle$-module
$V=\coker(p^I_1\otimes\id_{Q_0},\id_{P_0}\otimes q^1_1)$ and its $R$-free resolution
\beq\label{eq:res1}
\ldots \stackrel{q_0'}{\longrightarrow} Q_1'
\stackrel{q_1'}{\longrightarrow} Q_0'\stackrel{q_1'}{\longrightarrow} Q_1'
\stackrel{(p_1^I,q_1^1)}{\longrightarrow} P_0\otimes Q_0\rightarrow V \rightarrow 0\,,
\eeq
which turns into the matrix factorization $Q'$ after two steps. But now as an $R$-module
\beq
V=\coker(p_1^I,q_1^1)\cong\CC[x,y]/\langle \prod_{i\in I}(x-\eta_i y),y\rangle\cong
\CC[x]/\langle x^{|I|}\rangle=:\widetilde V\,,
\eeq
which has a two-periodic resolution 
\beq\label{eq:res2}
\ldots \stackrel{x^{d-|I|}}{\longrightarrow} R
\stackrel{x^{|I|}}{\longrightarrow} R\stackrel{x^{d-|I|}}{\longrightarrow} R
\stackrel{x^{|I|}}{\longrightarrow} R\rightarrow \widetilde{V} \rightarrow 0\,,
\eeq
corresponding to the matrix factorization $Q^{|I|}$. Hence, $Q'\cong Q^{|I|}$. 

Next we will show that also the boundary condition changing spectra between elementary boundary conditions specified by $Q^{\ell_1}$ and $Q^{\ell_2}$ can be induced from defect changing spectra of $P^{I_1}$ and $P^{I_2}$, $|I_i|=\ell_i$ upon fusion with $Q^1$. 

The spectrum between the
defects $P^{I_1}$ and $P^{I_2}$ depends very much on the divisibility properties of
$p_1^{I_1}, p_1^{I_2}$. In the extreme case $I_1=I_2$ the spectrum is purely bosonic,
whereas in the case that $I_1$ and $I_2$ have no common factors it is purely
fermionic. For our purposes, we are interested in having many fermionic defect 
changing fields, and hence we choose $I_1$ and $I_2$ such that the cardinality of their intersection
is minimized. If $\ell_1+\ell_2\leq d$, there are non-intersecting $I_1$ and $I_2$ with $|I_i|=\ell_i$. 
Then $p_0^{I_2}$ is always divisible by $p_1^{I_1}$.
If on the other hand
$\ell_1 + \ell_2 >d$ the intersection $I_1\cap I_2$ contains at least $\ell_1+\ell_2-d$ elements. If $I_1$ and $I_2$ are chosen to contain exactly $\ell_1+\ell_2-d$ elements, 
$p_1^{I_1}$ is divisible by $p_0^{I_2}$. 
This means that the condition for BRST-closedness of a defect changing field $T:P^{I_1}\rightarrow P^{I_2}$
\beq
p_0^{I_2} T_1 + T_0 p_1^{I_1}=0
\eeq
can be solved similarly to the case of the boundary conditions $Q^\ell$
\beqa
T_0 &=& -\frac{p_0^{I_2}}{p_1^{I_1}} T_1, \quad {\rm for} \ \ell_1 + \ell_2 \leq d, \\
T_1 &=& -\frac{p_1^{I_1}}{p_0^{I_2}} T_0, \quad {\rm for}\ \ell_1 + \ell_2 > d
\eeqa
BRST-exact fermions satisfy
\beqa
T_1 &=& \prod_{m\in I_1}(x -\eta_m y) \phi_1 + \phi_0 
\prod_{m\in I_2} (x-\eta_m y) \\
T_0 &=& \phi_0\prod_{m\in D\backslash I_1}(x -\eta_m y)  +  
\prod_{m\in D\backslash I_2} (x-\eta_m y) \phi_1\nonumber
\eeqa
Thus, the fermionic BRST-cohomology can be described as
\beq\label{eq:fermdefect}
T=(T_1,T_0)=(T_1,-{p_0^{I_2}\over p_1^{I_1}}T_1)\,,\quad
T_1\in\prod_{i\in I_1\cap I_2}(x-\eta_iy)\CC[x,y]/\langle p_1^{I_1},p_1^{I_2}\rangle\,.
\eeq
Next, we will show that upon fusion with $Q^1$ these defect changing spectra indeed induce
the boundary condition changing spectra between the respective $Q^{\ell_i}$.

To show that this is the case, we first determine the equivalence of the matrix factorizations 
$Q'=P^I\otimes Q^1$ and $Q^\ell$ with $\ell=|I|$. This can be easily done using the method described 
in Section \ref{fusiontools}. Namely, we just have to lift the isomorphism $V\cong\widetilde V$ to a map between the  resolutions \eq{eq:res1} and \eq{eq:res2}. Setting $\widehat R=\CC[x,y]/\langle W(x)\rangle$ we have to construct the $R$-module homomorphisms $r,r^*,r_i,r_i^*$ in 
\begin{equation}
\begin{array}{ccccccccccc}
\ldots
&\stackrel{q_1'}{\longrightarrow}
&\widehat{R}^2
&\stackrel{q_0'}{\longrightarrow}
&\widehat{R}^2
&\stackrel{(p_1^I,y)}{\longrightarrow}
&\widehat{R}
&\to
&\coker(p_1^I, y) 
&\to
& 0 \\
&
&\mapupdown{r_2}{r_2^*} 
& & \mapupdown{r_1}{r_1^*}
& & \mapupdown{r}{r^*}
& & \mapupdown{}{\cong}
& \\
\ldots
& \stackrel{x^\ell}{\longrightarrow}
& R
& \stackrel{x^{d-\ell}}{\longrightarrow}
& R
& \stackrel{x^\ell}{\longrightarrow}
& R
& \to
& \coker(x^{\ell})
& \to
& 0
\end{array}\ .
\end{equation}
The isomorphism can be lifted in the following way:
\beq
\begin{array}{ll}
r:\, (1\mapsto 1\,,\; y^i\mapsto 0)\,,&r^*\,: 1\mapsto 1\,,\\
r_1 =r \circ (1,\ 0) \,,&r_1^*= \left( \begin{array}{c} 1 \\  \frac{1}{y}(-p_1^I+x^{\ell}) \end{array} \right)\circ r^*\,,\\
r_2 =r\circ (1, \ 0) \,,  & r_2^* = \left( \begin{array}{c} 1 \\  \frac{1}{y}(-p_0^I+x^{d-\ell}) \end{array} \right)\circ r^*\,,\\
\ldots\;.&\end{array}
\eeq
Note that $p_1^I=x^{\ell} + y (\dots )$ and $p_0^I=x^{d-\ell}+y(\ldots )$ so that all the morphisms are well defined. As discussed in Section \ref{fusiontools} the morphisms $r_1,r_1^*,r_2,r_2^*$ indeed provide the equivalence of the matrix factorizations $Q'=P^I\otimes Q^1$ and $Q^\ell$, and they can be used to 
transfer defect changing fields $T:P^{I_1}\rightarrow P^{I_2}$ to boundary condition changing fields $t:Q^{\ell_1}\rightarrow Q^{\ell_2}$. Upon fusion with $Q^1$, a defect changing field $T$ is transferred
to a boundary condition changing field $T\otimes\id_{Q^1}$ on $Q'$. By means of the equivalence one obtains
\beqn\label{mmtransfer}
t_1&=&r_2^{(2)}\circ (T\otimes\id_{Q^1})_1\circ r_1^{(1)*}=r\circ T_1\circ r^*
\,,\\
t_0&=&r_1^{(2)}\circ (T\otimes\id_{Q^1})_0\circ r_2^{(1)*}=r\circ T_0\circ r^*\,,\nonumber
\eeqn
where $r_i^{(i)}, r_i^{(i)*}$ denote the equivalences of $P^{I_i}\otimes Q^1\cong Q^{\ell_i}$ respectively.
Thus, the $t_i$ are obtained from the $T_i$ by setting $y=0$. Comparing the fermionic spectra \eq{eq:fermdefect} of defect changing fields $P^{I_1}\rightarrow P^{I_2}$ and the ones \eq{eq:fermboundary} of boundary condition changing fields $Q^{\ell_1}\rightarrow Q^{\ell_2}$, one finds that the entire fermionic boundary spectra
can be induced by defect changing fields upon fusion with $Q^1$. Thus, for minimal models, all boundary RG flows can be pulled back to the bulk using defects.

\subsection{Example: Minimal model orbifolds}\label{orbifold}

As a next example, we consider the $\ZZ_d$-orbifold of the Landau-Ginzburg model with superpotential
$W=x^d$, where the orbifold group acts on the chiral superfield $x$ by phase multiplication. In fact, these orbifold theories are mirror to the original unorbifolded Landau-Ginzburg models, and B-type defects and boundary conditions in the orbifolds correspond to A-type defects and boundary conditions in the unorbifolded theories. As mentioned in Section \ref{sec:MF}, B-type boundary conditions and matrix factorizations in Landau-Ginzburg orbifolds are represented by equivariant matrix factorizations of the
respective superpotentials. 
For instance the $\ZZ_d$-equivariant rank one factorizations of $W$ are given by
\beq
Q^{\ell}_m:\CC[x][m+\ell] \maplr{x^\ell}{x^{d-\ell}} \CC[x] [m] \,,
\eeq
where the $\ZZ_d$-representation on $\CC[x][m]$ is specified by the action of the generator on $1\in\CC[x]$: $1\mapsto e^{2\pi i m\over d} 1$. Also in the orbifold models all matrix factorizations can be obtained by cones of such rank-one factorizations. 

B-type defects in the orbifold models are represented by $\Gamma=\ZZ_d\times\ZZ_d$-equivariant matrix factorizations of $W(x)-W(y)$, where the first $\ZZ_d$ acts on $x$ only, and the second one on $y$. 
Indeed, by means of the orbifold construction, one can obtain such factorizations out of the non-equivariant $P^I$ defined in equation \eq{idartig}. Roughly speaking, one chooses a representation
of the stabilizer subgroup $\Gamma_{\rm stab}\subset\Gamma$ under which $P^I$ is invariant and then takes its $\Gamma/\Gamma_{\rm stab}$-orbit. In this case $\Gamma_{\rm stab}$ is the diagonal $\ZZ_d$-subgroup, so one obtains a sum 
\beq\label{diagform}
P^I_m=\bigoplus_{\gamma\in\ZZ_d} P^{\gamma(I)}[m]\,,
\eeq
where the re\-pre\-sen\-tation of the diagonal $\ZZ_d$-subgroup is indicated by $(\cdot)[m]$ and $$\gamma(\{i_1,\ldots,i_r\})=\{i_1+\gamma\,{\rm mod}\, d,\ldots,i_r+\gamma\,{\rm mod}\, d\}$$ is the cyclic shift of $I$.
This is a diagonal $d$-dimensional matrix factorization, on which the action of $\Gamma$ is non-diagonal however. But it can be diagonalized. Denoting the basis in which the matrix factorization is diagonal by $e_i$, we can diagonalize the $\Gamma$-action by the change of basis
\beq\label{basechange1}
\hat{e}_n = \sum_{i=1}^d \eta^{in} e_i
\eeq
which is inverted by
\beq\label{basechange2}
e_m=\frac{1}{d} \sum_i \eta^{mi} \hat{e}_i.
\eeq
In the basis $\hat{e}_n$, $P^I_M$ takes the form
\beq
P_m^I:
\CC[x,y]^d \left( \begin{array}{c} 
\s \left[m+|I|,0\right] \\  \s \left[m+1+|I|, -1\right] \\ \vdots \\ \s \left[m+(d-1)+|I|, -(d-1) \right] \end{array} \right) 
\maplr{\!\!\!\!\!\!\!\!\!p_1^I(X,Y)}{\!\!\!\!\!\!\!\!\!p_0^I(X,Y)}
 \CC[x,y]^d \left( \begin{array}{c} 
 \scriptstyle \left[m,0 \right] \\ \s \left[m+1, -1\right] \\ \vdots \\ \s \left[m+(d-1), -(d-1) \right] \end{array} \right) \,,
 \eeq
where $X$ and $Y$ are the $d\times d$-matrices $X=x 1_d$ and $Y=y\Xi$, with 
$\Xi$ the $d$-dimensional shift matrix
\beq
\Xi_{ab}=\delta_{a,b+1}^{(d)}\,.
\eeq
Note that because of the orbit formation in the orbifold construction not all of the $P_m^I$ are inequivalent. In fact, $P_m^I\cong P_m^J$ if $J$ is a (cyclic) shift of $I$, \ie $J=I+n\,{\rm mod}\, d$. 

The fusion of $P_m^I$ with $Q^1_0$ can be easily calculated. Using the method already employed in Section \ref{sec:mmdefects}, one obtains
\beq\label{unorbfus}
P_m^I*Q^1_0\cong \bigoplus_{i\in\ZZ_d} Q^{|I|}_{m+i}[-i]\,,
\eeq
where here $\cdot[-i]$ denotes the representation of the second $\ZZ_d$, the orbifold group of the model
squeezed in between defect and boundary. The fusion in the orbifold model is the $\Gamma_{\rm squeezed}$-invariant part of \eq{unorbfus}
\beq
P_m^I*_{\rm orb} Q^1_0\cong Q^{|I|}_m\,.
\eeq
Hence, also in the orbifold theory, one can generate all elementary matrix factorizations $Q_m^\ell$ by 
fusing defect factorizations $P_m^I$ with $Q^1_0$.

Indeed, there is another way to obtain the fusion \eq{unorbfus}. Namely, we one can use
the diagonal form \eq{diagform} $P^I_m$, which is a direct sum of the ordinary rank-one factorizations $P^I$. In this way, one reduces the problem to the problem in the non-orbifolded situation discussed in Section \ref{sec:mmdefects}. Since the result of the fusion of
$P^I$ with $Q^1$ only depends on the cardinality of $I$, the fusion $P_m^I*Q^1_0$ just gives a direct sum of $d$ equal summands $Q^{|I|}$. To bring this in the basis in which the $\Gamma$-action is diagonal, we have to do the change of basis (\ref{basechange1},\ref{basechange2}). Being diagonal, this does not change the result however, and we arrive at \eq{unorbfus}.

Having established that one can generate all $Q^{\ell}_m$ by fusing defects $P^I_m$ with $Q^1_0$, we 
would like to show next that also the entire spectra of boundary operators $Q^{\ell_1}_{m_1}\rightarrow Q^{\ell_2}_{m_2}$ can be induced upon fusion with $Q^1_0$ by defect changing operators $P^{I_1}_{m_1}\rightarrow P^{I_2}_{m_2}$ with $|I_i|=\ell_i$. 

As in the unorbifolded theory we choose $I_1$, $I_2$ in such a way that $|I_1\cap I_2|$ is minimized.
Indeed, it is easy to see that for a given $T(x,y)\in\HH^1(P^{I_1},P^{I_2})$ in the unorbifolded theory of degree $\ell_1+m_1-m_2\,{\rm mod}\,d$ for $T_1$ and degree $\ell_2+m_2-m_1$ for $T_0$, 
$T(X,Y)\in\HH^1_{\rm orb}(P^{I_1}_{m_1},P^{I_2}_{m_2})$. 
To transfer this boundary condition changing field to the fused boundary we make use of the diagonal form \eq{diagform}. This allows us to reduce the problem to the unorbifolded problem. Using the equivalences in the unorbifolded case \eq{mmtransfer} and then projecting to the invariant part of the fusion product, we obtain
\beq
t=r\circ T\circ r^*=T(x,0)
\eeq
for the boundary changing field. As in the unorbifolded case, comparing defect and boundary BRST-cohomology we find that before the orbifold projection we obtain the entire boundary spectra this way. But of course also the projections agree. Hence, we arrive at the conclusion that in the orbifold models as well
the entire boundary changing spectra can be induced from defect changing fields 
by fusion of defects $P^I_m$ with $Q^1_0$.

\subsection{Example: Tensor products of identical LG models}

Another simple example is the tensor product of two identical minimal models. It turns out that this is not any simpler than the more general
case of a product of two arbitrary identical Landau-Ginzburg models. In fact, for ease of notation
 we will consider tensor products of Landau-Ginzburg models with their conjugates in the following.
 That means the models have 
 chiral superfields $x_1,\ldots, x_N,\widehat{x}_1,\ldots,\widehat{x}_N$ and superpotential $W(x_i)-W(\widehat{x}_i)$ instead of $W(x_i)+W(\widehat{x}_i)$. The construction below easily carries over to the tensor product of identical Landau-Ginzburg models. 

Now, given any matrix factorization $Q(x_i,\widehat{x}_i)$ of $W(x_i)-W(\widehat{x}_i)$, we define
the defect matrix factorization of $W(x_i)-W(\widehat{x}_i)-W(y_i)+W(\widehat{y}_i)$ as the tensor product
\beq
P_Q:=Q(x_i,y_i)\otimes\Id(\widehat{x}_i,\widehat{y}_i)\,,
\eeq
where $\Id(\widehat{x}_i,\widehat{y}_i)$ is the matrix factorization representing the identity defect in the second tensor factor.

Of course, fusing $P_Q$ with the identity matrix factorization 
\beq
E=\Id(y_i,\widehat{y}_i)
\eeq
between the two tensor factors gives back the matrix factorization $Q$:
\beq
P_Q*E\cong Q\,.
\eeq
Here one only needs to make repeated use of the fact that $\Id*P\cong P$ for any matrix factorization $P$. Hence, in these models every matrix factorization $Q$ can be obtained by fusing a defect matrix factorization $P_Q$ with a fixed matrix factorization $E$. This is true in particular for families $Q(t)$ of matrix factorizations.
Therefore, all perturbations of boundary conditions can be pulled back into the bulk by means of the defects $P_Q$ in these models.

Of course it is also clear from the above that the defect changing fields
\beq
\HH^1(Q^1,Q^2)\otimes\id_{\Id}\subset \HH^1(P_{Q^1},P_{Q^2})
\eeq
induce the respective boundary condition changing fields between the fused boundaries
$P_{Q^i}*E\cong Q^i$. 

In fact, the arguments used to arrive at this conclusion do not depend on the matrix factorization formalism, so we expect that the result carries over to tensor products of arbitrary $N=(2,2)$-theories with their conjugates. 

\subsection{Defect induced boundary flows in CFT}

The fusion of perturbed defects has been considered on the level of the full
conformal field theory for rational models with diagonal modular invariants 
in \cite{Runkel:2007wd}. 
There, only defects which preserve both, the holomorphic and antiholomorphic 
$W$-algebras on the full complex plane were considered. These defects in particular 
preserve both copies of the Virasoro algebra and are therefore topological, which implies
that their fusion is non-singular.

To ensure that also the perturbed defects (taken here to extend parallel to the real line)
can be moved smoothly along the imaginary axis, the perturbations are restricted to chiral defect fields $\phi(z)$, $\frac{\partial}{\partial\bar{z}} \phi(z) =0$. Defects perturbed by chiral fields still commute with the Hamiltonian generating translations along the imaginary axis, and hence can be fused smoothly with
parallel defects. Moreover, the perturbations are further restricted in \cite{Runkel:2007wd} by demanding that only fields in a single fixed representation occur.

The result of the fusion of two defects perturbed in this way is obtained as a bunch of defects resulting from the fusion of the unperturbed defects perturbed again by defect changing fields in the fixed representation.

For rational CFTs with charge conjugate modular invariant,
defect operators corresponding to topological defects 
can immediately be written down \cite{Petkova:2000ip}
\beq
D_{J} = \sum_j \frac{S_{Jj}}{S_{0j}} {\rm P}_{j\bar{j}} \ .
\eeq
Here, $J,j$ specify irreducible representations of the chiral symmetry algebra 
and take values in some index set ${\cal I}$. ${\rm P}_{j \bar{j}}$ are projection operators on the representation spaces ${\cal V}_j \otimes {\cal V}_{\bar{j}}$, and $S$ denotes the modular $S$-matrix
of the respective characters $\chi_j$. Fusion of these defects can be obtained by composing the respective operators, and using the Verlinde formula, one easily obtains
\beq
D_J * D_{J'} = \sum_{J''} {\cal N}_{J J'}^{J''} D_{J''},
\eeq
where ${\cal N}$ denote the fusion rule coefficients.

Likewise, boundary states for symmetry preserving boundary conditions 
are given by Cardy's formula
\beq
\kket{J} = \sum_{j} \frac{S_{Jj}}{\sqrt{S_{0j}}} |{j}\rangle\!\rangle \,,
\eeq
where $|j\rangle\!\rangle$ denote the Ishibashi states in the sector $\VV_j\otimes\overline{\VV}_{\bar j}$. 

Fusion of topological defects with boundary conditions can be calculated by applying the respective
defect operators to the boundary states. For the defects and boundary states above this yields
\beq
D_J  \kket{J'} =  \sum_{J''} {\cal N}_{J J'}^{J''} \kket{J''} \ .
\eeq
This implies in particular that all Cardy boundary conditions $\kket{J}$ 
can be
obtained by fusing the topological defects $D_J$ with the boundary condition 
$\kket{0}$ associated to the vacuum representation.
This is very much like in the case of B-type boundary conditions in Landau-Ginzburg models
discussed in Sections \ref{sec:mmbranes} and \ref{orbifold} above, where all boundary conditions
can be produced by fusing defects with a fixed linear matrix factorization.

The spectra of defect changing fields between the defects $D_J$ and the spectrum of boundary 
condition changing fields between the Cardy boundary conditions $\kket{J}$ can be easily determined 
in the RCFT setting
\beqn\label{defhilb}
{\cal H}^D_{J J'}&=& \bigoplus_{j,j',j''} {\cal N}_{j j'}^{j''} {\cal N}_{j'' J'}^J {\cal V}_j \otimes {\overline{\cal V}}_{j'} \,,\\
{\cal H}^B_{J J'}&=& \bigoplus_{j} {\cal N}_{J' j }^{J}  {\cal V}_j  \ .
\eeqn
The chiral defect changing fields are obtained by setting $j'=0$ in \eq{defhilb}. One immediately finds
that the space of chiral defect changing fields between defects $D_J$ and $D_{J'}$ is isomorphic to the
space of boundary condition changing fields between $\kket{J}$ and $\kket{J'}$.

This implies that indeed all boundary perturbations
  can be obtained by fusing the boundary condition
$\kket{0}$ with chirally perturbed defects. 

For instance a boundary flow of a sum of boundary conditions $\kket{J}\oplus\kket{J'}$ generated by  a boundary condition changing field in representation $j$ can be obtained by fusing boundary condition
$\kket{0}$ with a defect $D_J\oplus D_{J'}$ perturbed by the chiral defect changing field in that same representation\footnote{Representations of fields are not changed when they are transferred 
in the fusion process.}.

On a formal level this is quite similar to what we have seen for matrix
factorizations: All boundary conditions can be generated by fusing defects with a particular boundary condition, and all boundary condition changing fields  can be induced by choosing particular defect changing fields.


To make the relation completely precise, one can consider
the example of the supersymmetric minimal model with A-type boundary conditions and defects, or equivalently B-type boundary conditions and defects in the orbifold of the minimal model.
Here both a description in terms of matrix factorizations
and in terms of rational conformal field theory is available.
On the level of matrix factorizations this is the Landau-Ginzburg orbifold discussed in Section \ref{orbifold}.

The $N=(2,2)$-superconformal minimal models ${\mathcal M}_k$ are rational with respect to the $N=2$ super
Virasoro algebra at central charge $c_k={3k\over k+2}$. In fact, the
bosonic part of this algebra can be realized as the coset W-algebra
\beq\label{coset}
\left({\rm SVir}_{c_k}\right)_{\rm bos} = 
{\widehat{\mathfrak{su}}(2)_k\oplus\widehat{\mathfrak{u}}(1)_4\over \widehat{\mathfrak{u}}(1)_{2k+4}}\,,
\eeq  
and the respective coset CFT can be obtained from ${\mathcal M}_k$ by
a non-chiral GSO projection. 

The Hilbert space $\HH^k$ of ${\mathcal M}_k$ decomposes into irreducible
highest weight representations of holomorphic and antiholomorphic
super Virasoro algebras, but it is convenient to decompose it further
into irreducible highest weight representations $\VV_{[l,m,s]}$ of the
bosonic subalgebra \eq{coset}. These representations are labelled by 
\beq
[l,m,s]\in\mathcal{I}_k:=\{(l,m,s)\,|\,0\leq l\leq
k,\,m\in\ZZ_{2k+4},\,s\in\ZZ_4,\,l+m+s\in 2\ZZ\}/\sim\,,
\eeq
where $[l,m,s]\sim[k-l,m+k+2,s+2]$ is the field identification. The highest weight representation
of the full super Virasoro algebra are given by
\beq
\VV_{[l,m]}:=\VV_{[l,m,(l+m)\,{\rm
    mod}\,2]}\oplus\VV_{[l,m,(l+m)\,{\rm mod}\,2+2]}\,.
\eeq
For $(l+m)$ even $\VV_{[l,m]}$ is in the NS-, for $(l+m)$ odd in the R-sector.
Here $[l,m]\in{\mathcal J}_k:=\{(l,m)\,|\,0\leq l\leq
k,\,m\in\ZZ_{2k+4}\}/\sim$, $[l,m]\sim[k-l,m+k+2]$. 
The Hilbert spaces of ${\mathcal
M}_k$ in the NSNS- and RR-sectors then read
\beq
\HH^k_{NSNS} 
\cong\bigoplus_{\stackrel{[l,m]\in{\mathcal J}_k}{l+m\,{\rm
      even}}}\VV_{[l,m]}\otimes\overline{\VV}_{[l,m]} \,,\qquad
\HH^k_{RR}
\cong\bigoplus_{\stackrel{[l,m]\in{\mathcal J}_k}{l+m\,{\rm
      odd}}}\VV_{[l,m]}\otimes\overline{\VV}_{[l,m]} \,.
\eeq
To obtain a CFT with a modular invariant partition function from this fully
supersymmetric theory, one needs to perform a GSO projection. In the case at hand, there are two possibilities, a type 0A and a type 0B projection, distinguished by the action of $(-1)^F$.  We will consider the type 0B case, where
states in the sector $\VV_{[l,m,s]} \otimes \VV_{[l,m,-s]}$ are invariant under
the projection.

The defects of the theory with either GSO projection have been given in \cite{Brunner:2007qu}.
The general form of the defect operators in the Cardy case is
\beq
{\mathcal D}=\sum_{\stackrel{[l,m,s],\bar s}{s-\bar s\,{\rm
      even}}}{\mathcal D}^{[l,m,s,\bar s]} {\rm P}_{[l,m,s,\bar s]}\,,
\eeq
where ${\rm P}_{[l,m,s,\bar s]}$ is a projector on the modules
$\VV_{[l,m,s]}\otimes\overline{\VV}_{[l,m,\bar s]}$ of the bosonic subalgebra.
The solutions for the coefficients are given by
\beq\label{qdim}
{\mathcal D}_{[ L, M,
S,\bar S]}^{[l,m,s,\bar s]}=e^{-i\pi {\bar S(s+\bar s)\over 2}}{S_{[ L, M,
  S-\bar S][l,m,s]}\over
S_{[0,0,0],[l,m,s]}}\,,
\eeq
where the different defects are specified by $[ L,
M, S,\bar S]$ with $[L,M,S-\bar S]\in {\mathcal I}_k$, and 
\beq
S_{[L,M,S][l,m,s]}={1\over k+2}e^{-i\pi{Ss\over 2}}e^{i\pi{Mm\over
k+2}}\sin\left(\pi{(L+1)(l+1)\over k+2}\right)
\eeq
is the modular S-matrix for the coset representations
$\VV_{[l,m,s]}$.

In the orbifold theory, modding out the $\ZZ_{k+2}$ 
phase symmetry acting on the ${\mathfrak u}_{2k+4}$ labels projects the Hilbert space of ${\cal M}_k$ on the subsector with $m=0$. Together with the
twisted sectors, the new Hilbert space takes the form
\beq
\HH^k_{NSNS} 
\cong\bigoplus_{\stackrel{[l,m]\in{\mathcal J}_k}{l+m\,{\rm
      even}}}\VV_{[l,m]}\otimes\overline{\VV}_{[l,-m]} \,,\qquad
\HH^k_{RR}
\cong\bigoplus_{\stackrel{[l,m]\in{\mathcal J}_k}{l+m\,{\rm
      odd}}}\VV_{[l,m]}\otimes\overline{\VV}_{[l,-m]} \,.
\eeq
The defect operators of the orbifold theory look very similar to that
of the original theory:
\beq\label{orbifold_defect}
{\mathcal D}^{\rm orb}=\sum_{\stackrel{[l,m,s],\bar s}{s-\bar s\,{\rm
      even}}}{\mathcal D}^{[l,m,s,\bar s]} {\rm P}^{\rm orb}_{[l,m,s,\bar s]}\,,
\eeq
where now  ${\rm P}^{\rm orb}_{[l,m,s,\bar s]}$ is a projector on ${\cal V}_{[l,m,s]} \otimes {\cal V}_{[l,-m,\bar{s}]}$. The coefficients are given by (\ref{qdim})
just like in the unorbifolded case. 

For this reason, also the fusion algebra between defects in the orbifold theory is the same as the one 
in the original unorbifolded model \cite{Brunner:2007qu}
\beq\label{CFTdefcomp}
{\mathcal D}^{\rm orb}_{[ L_1, M_1, S_1,\bar S_1]}{\mathcal D}^{\rm orb}_{[ L_2,
M_2, S_2,\bar S_2]}=\sum_{ L}{\mathcal N}_{ L_1  L_2}^{ L}{\mathcal D}^{\rm orb}_{[ L,
M_1+ M_2, S_1+ S_2,\bar S_1+\bar S_2]}\,.
\eeq
For $L=0$ these defects are group like. For the original theory, the defects ${\mathcal D}_{[0,M,0,0]}$ realize the action of the orbifold group, whereas in the orbifold theory,  the ${\mathcal D}^{\rm orb}_{[0,M,0,0]}$ realize the
corresponding quantum symmetry.

For later use, we calculate the defect changing spectrum. For this, we use the
folding trick and map the defects to  permutation boundary states in the doubled theory. These boundary states have been analyzed in detail for the unorbifolded case \cite{Brunner:2005fv,Enger:2005jk}. To summarize, in the original, unorbifolded theory, the permutation
B type boundary states are given by
\beqa\label{bdstates}
&&|\!|  [ L,M,S_1,S_2] \rangle\!\rangle\label{permstates}\\
& &\!=\! \frac{1}{2\, \sqrt{2}}\! \sum_{l,m,s_1,s_2}\! 
\frac{S_{Ll}}{S_{0l}} \, e^{i\pi M m / (k+2)}\, 
e^{-i\pi (S_1 s_1 - S_2 s_2)/2}\,
|[l,m,s_1]\otimes [l,-m,-s_2]\rangle\!\rangle^\sigma  
\,,\nonumber 
\eeqa
where the sum runs over all $l,m,s_1$ and $s_2$ for which 
\beq
l+m+s_1 \quad \hbox{and} \quad s_1 -s_2\quad \hbox{are even.}
\eeq
Here, $|[l,m,s_1]\otimes [l,-m,-s_2]\rangle\!\rangle^\sigma$ are B-type 
permutation Ishibashi states in the sectors
\beq
\Bigl(\V_{[l,m,s_1]}\otimes \V_{[l,-m,-s_2]}\Bigr) \otimes
\Bigl(\bar\V_{[l,m,s_2]}\otimes \bar\V_{[l,-m,-s_1]}\Bigr) \,,
\eeq
which means that they intertwine the respective supersymmetry algebras of the two tensor factors. 
In the orbifold theory, the boundary states are similarly given by
\begin{eqnarray}\label{orbbdstates}
&& \kket{ [  L,M,S_1,S_2]}_{\rm orb} \\
& & = \frac{1}{2 \sqrt{2}} \sum_{l,m,s_1,s_2} 
\frac{S_{Ll}}{S_{0l}} \, e^{i\pi M m / (k+2)}\, 
e^{-i\pi (S_1 s_1 - S_2 s_2)/2}\,
 |[l,m,s_1]\otimes [l,m,-s_2]\rangle\!\rangle^{\sigma}_{{\rm orb}}  
\,,\nonumber 
\end{eqnarray}
where now the permutation Ishibashi states with $m\neq 0$ come from the twisted sectors
\beq
\Bigl(\V_{[l,m,s_1]}\otimes \V_{[l,m,-s_2]}\Bigr) \otimes
\Bigl(\bar\V_{[l,m,s_2]}\otimes \bar\V_{[l,m,-s_1]}\Bigr) \,.
\eeq
These boundary states can be obtained directly from the defect (\ref{orbifold_defect}) by means of the folding trick. It can also be obtained from the un-orbifolded boundary states \eq{bdstates} using the orbifold construction.
To see this, note that the boundary states are invariant under the diagonal $\ZZ_{k+2} \subset \ZZ_{k+2} \times \ZZ_{k+2}$, leading to resolved boundary states labelled by an additional $\ZZ_{k+2}$-representation label $M'$, which specifies the representation in the boundary sectors. Orbifolding by the second $\ZZ_{k+2}$-factor introduces an orbit of boundary states of different $M$. The result is \eq{orbbdstates}.
In our notation we do not distinguish between labels $M$ and $M'$ although from the orbifold point of view these labels play different roles.

Note that the permutation boundary states of the two theories differs only in a minus sign in front of one of the $m$ labels in the Ishibashi states. As a consequence, also the one-loop amplitudes are almost 
identical to the ones 
\beqn
&&  \langle\!\langle  [L,M,S_1,S_2] |\!|  
q^{\frac{1}{2}(L_0 + \bar{L}_0) - \frac{c}{12}}
 |\!|  [\hat{L},\hat{M},\hat{S}_1,\hat{S}_2] \rangle\!\rangle   
= \!\! \sum_{[l_i',m_i',s_i']}\!\!  
\chi_{[l_1',m_1',s_1']}(\tilde{q})\, 
\chi_{[l_2',m_2',s_2']}(\tilde{q})\nonumber
\\
& &  \sum_{\hat{l}}
\Bigl[ {\cal N}_{\hat{l} \hat{L}}{}^{L} \, {\cal N}_{l_1' l_2'}{}^{\hat{l}}\,
\delta^{(2k+4)}(\Delta M+m'_1-m'_2) \nonumber \\
& & \quad \times
\Bigl( \delta^{(4)}(\Delta S_1+s_1')\, \delta^{(4)}(\Delta S_2+s_2')
+ \delta^{(4)}(\Delta S_1 +2+s_1')\, \delta^{(4)}(\Delta S_2 +2+s_2') 
\Bigr) \nonumber \\
& &   + {\cal N}_{\hat{l}\, k-\hat{L}}{}^{L} \, {\cal N}_{l_1' l_2'}{}^{\hat{l}}\,
\delta^{(2k+4)}(\Delta M+k+2+m'_1-m'_2) \nonumber \\
& &  \quad \times 
\Bigl( \delta^{(4)}(\Delta S_1+2+s_1')\, \delta^{(4)}(\Delta S_2+s_2') 
+ \delta^{(4)}(\Delta S_1+s_1')\, \delta^{(4)}(\Delta S_2+2+s_2') 
\Bigr) \Bigr] \,, \nonumber
\eeqn
of the original unorbifolded theory.
Here  $\Delta M = \hat{M} - M$ and $\Delta S_i = \hat{S}_i-S_i$. 
In particular, we find that the boundary spectrum of $\kket{[0,0,0,0]}$ is isomorphic to the bulk spectrum,
which is expected, because it is isomorphic to the spectrum of defect fields on the trivial defect.

In the orbifold theory one obtains
\beqa\label{spectrum}
&& \langle\!\langle  [L,M,S_1,S_2] |\!|  
q^{\frac{1}{2}(L_0 + \bar{L}_0) - \frac{c}{12}}
 |\!|  [\hat{L},\hat{M},\hat{S}_1,\hat{S}_2] \rangle\!\rangle_{\rm orb}   
\!\!= \!\!\!\! \sum_{[l_i',m_i',s_i']}\!\!\!  
\chi_{[l_1',m_1',s_1']}(\tilde{q})\, 
\chi_{[l_2',m_2',s_2']}(\tilde{q}) \nonumber
\\
& & \sum_{\hat{l}}
\Bigl[ {\cal N}_{\hat{l} \hat{L}}{}^{L} \, {\cal N}_{l_1' l_2'}{}^{\hat{l}}\,
\delta^{(2k+4)}(\Delta M+m'_1+m'_2) \nonumber \\
& & \quad \times
\Bigl( \delta^{(4)}(\Delta S_1+s_1')\, \delta^{(4)}(\Delta S_2+s_2')
+ \delta^{(4)}(\Delta S_1 +2+s_1')\, \delta^{(4)}(\Delta S_2 +2+s_2') 
\Bigr) \nonumber \\
& &  + {\cal N}_{\hat{l}\, k-\hat{L}}{}^{L} \, {\cal N}_{l_1' l_2'}{}^{\hat{l}}\,
\delta^{(2k+4)}(\Delta M+k+2+m'_1+m'_2) \nonumber \\
& & \quad \times 
\Bigl( \delta^{(4)}(\Delta S_1+2+s_1')\, \delta^{(4)}(\Delta S_2+s_2') 
+ \delta^{(4)}(\Delta S_1+s_1')\, \delta^{(4)}(\Delta S_2+2+s_2') 
\Bigr) \Bigr] \,, \nonumber
\eeqa
where as before $\Delta M = \hat{M} - M$ and $\Delta S_i = \hat{S}_i-S_i$. 
As alluded to above, the only difference between the open string
spectra for the orbifold and the original theory is the sign with which $m_2$ enters. The reason for this is of course that in the bulk of the orbifold theory ${\cal V}_{[l,m,s]}$
is paired with $\bar{{\cal V}}_{[l,-m,s]}$ instead of $\bar{{\cal V}}_{[l,m,s]}$ so that the
B-type permutation Ishibashi states are from a conjugate sector compared to the original theory. A modular transformation to the open string sector then leads to a sign flip for $m_2'$, which is the only difference. 

Let us now discuss the B-type boundary states in minimal models and their 
spectra. For the original unorbifolded model, the boundary states
are given by
\beq\label{bs-original}
|\!| L,S \rangle\!\rangle = 
\sqrt{k+2} \,
\sum_{l+s\in 2\ZZ} 
\frac{S_{[L,0,S],[l,0,s]}}{\sqrt{S_{[l,0,s],[0,0,0]}}} \, 
| [l,0,s]\rangle\!\rangle \,.
\eeq
These boundary states are not of Cardy type, but require an additional projection on Ishibashi states that satisfy B-type boundary conditions.

This is different in the orbifold theory, where indeed the standard Cardy construction can be applied 
to B-type boundary conditions.
The Ishibashi states $|[l,m,s]\rangle\!\rangle_{\rm orb}$ are from the sectors
${\cal V}_{[l,m,s]} \otimes \bar{{\cal V}}_{[l,-m,-s]}$ 
and the boundary states are explicitly given by
\beq
\kket{[L,M,S]}_{\rm orb} = \sum_{[l,m,s]} \frac{S_{[L,M,S][l,m,s]}}{\sqrt{S_{[0,0,0][l,m,s]}}} |[l,m,s]\rangle\!\rangle_{\rm orb} \, .
\eeq
These boundary states can of course also be obtained from the states \eq{bs-original} by the orbifold construction. The spectrum of boundary condition changing fields between two such orbifold boundary conditions is given by the partition function
\beqa\nonumber
&& \langle \! \langle  [L,M,S]|q^{\frac{1}{2}(L_0 + \bar{L}_0)-\frac{c}{24}}\kket{[L',M',S']}_{\rm orb}\\ \label{brane_pf}
&& ~~~ = \sum_{[l,m,s]} \big(N_{LL'}^l \delta^{(4)}(S'-S+s) \delta^{(2k+4)}(M'-M+m)
\\ \nonumber
&& ~~~+N_{LL'}^{k-l} \delta^{(4)}(S'-S+s+2) \delta^{(2k+4)}(M'-M+m+k+2) \big) \chi_{[l,m,s]}(q)
\eeqa

Being a special case of an RCFT with diagonal modular invariant, all supersymmetry preserving boundary flows between B-type boundary conditions in the orbifold theory should be generated by fusion of chirally perturbed B-type topological defects with the boundary condition $\kket{[0,0,0]}_{\rm orb}$. 
Indeed, fusion of the defects ${\cal D}^{\rm orb}_{[L_1,M_1,S_1,\bar{S}_1]}$ and boundary conditions 
$\kket{[L_2,M_2,S_2]}$ is given by
\beq
{\mathcal D}^{\rm orb}_{[L_1,M_1,S_1,\bar{S}_1]} \kket{[L_2,M_2,S_2]}_{\rm orb}
=\sum_L {\mathcal N}_{L_1 L_2}^L \kket{L,M_1+M_2, S_1-\bar{S}_1+ S_2}_{\rm orb} \, .
\eeq
In particular ${\cal D}^{\rm orb}_{[L,M,S,0]}\kket{[0,0,0]}_{\rm orb}=\kket{[L,M,S]}_{\rm orb}$. Hence, all
boundary conditions can be obtained by fusing boundary condition $\kket{[0,0,0]}_{\rm orb}$ with defects ${\cal D}^{\rm orb}_{[L,M,S,0]}$. Moreover, 
the chiral defect changing spectrum between defects ${\cal D}^{\rm orb}_{[L_1,M_1,S_1,0]}$ and ${\cal D}^{\rm orb}_{[L_2,M_2,S_2,0]}$, which can be obtained from the spectrum \eq{spectrum} of boundary condition changing operators of permutation boundary conditions in the folded model by setting $[l_2',m_2',s_2']=[0,0,0]$ is isomorphic to the spectrum of boundary condition changing operators between boundary conditions $\kket{[L_1,M_1,S_1]}_{\rm orb}$ and $\kket{[L_2,M_2,S_2]}_{\rm orb}$. 

Thus, in orbifolds of minimal models, we have explicitly seen that perturbations of B-type supersymmetric boundary condition can be pulled back into the bulk by chirally perturbed topological defects.

This can be compared to our discussion of boundary flows in Landau-Ginzburg orbifolds  in Section \ref{orbifold}. Namely, the minimal model ${\cal M}_{d-2}$ is the IR fixed point of a Landau-Ginzburg model
with superpotential $W=x^d$, and the same is true for the $\ZZ_d$-orbifolds of the respective models. 

Thus, B-type boundary conditions and defects in the (orbifold of the) minimal model ${\cal M}_{d-2}$ can nicely be represented by (equivariant) matrix factorizations. For the minimal model one obtains\footnote{Note that the matrix factorizations only describe B-type boundary states and defects with the same spin structure, which we chose by setting all the $S$-labels to zero.}
\beq
\begin{array}{ccc}
Q^\ell&\leftrightarrow& \kket{\ell-1,0}\\
P^{\{m,m+1,\ldots,m+\ell\}} & \leftrightarrow & {\mathcal D}_{[\ell,\ell+2m,0,0]}
\end{array}
\,.
\eeq
Note that there are more defect matrix factorizations than there are topological defects in the CFT. Namely, only those matrix factorizations $P^I$ 
have an interpretation as topological defects in the CFT for which $I$ is a set of consecutive integers (${\rm mod}\, d$) \cite{Brunner:2005fv,Enger:2005jk}.

In the orbifold theory, the relation is 
\beq
\begin{array}{ccc}
Q^\ell_m&\leftrightarrow& \kket{[\ell-1,\ell-1-2m,0]}\\
P^{\{m,m+1,\ldots,m+\ell\}}_M & \leftrightarrow & {\mathcal D}_{[\ell,\ell-2M,0,0]}
\end{array}
\,.
\eeq
Note that in the orbifold model $P_M^{\{m,m+1,\ldots,m+l\}}\cong P_M^{\{m',m'+1,\ldots,m'+l\}}$.

Comparing the discussion of B-type boundary flows in the matrix factorization approach and the full CFT we find complete agreement. The boundary condition $\kket{[0,0,0]}_{\rm orb}$ out of which all boundary conditions can be generated by means of fusion with topological defects corresponds to the matrix factorization $Q^1_0$, which was used in the same way in the matrix factorization approach. 
Of course, also the defects which are used for this purpose coincide, when we restrict to $P^I_M$ with
index sets consisting of consecutive (${\rm mod}\,d$) integers. Finally, inspection of the defect changing spectra in the CFT and the matrix factorization approach shows that we have indeed chosen the same defect changing fields to induce boundary condition changing fields in the fusion with $\kket{[0,0,0]}_{\rm orb}$ and $Q^1_0$ respectively.

\section{Braid group actions and defects}\label{sec:braidgroup}
In string theory, actions of braid groups on D-brane categories arise
in various contexts. For example one finds braid group actions on A-type branes whenever the target space manifold contains an $A_{m}$ chain of Lagrangian spheres ${\cal L}_i$, which have intersections
\beq\label{A_n}
({\cal L}_i \cap {\cal L}_j) = \left\{ \begin{array}{cc} 1 & |i-j|=1 \\ 0 & |i-j| > 1 \end{array} \right.
\eeq
In particular, such configurations arise when the compactification manifold degenerates into a singular space with singularity of type $A_{m}$. On the level of the homology it is well-known that probe cycles undergo a Picard-Lefschetz monodromy transformation when encircling a locus where one of the spheres shrinks to zero size. This transformation acts as
\beq\label{lefschetz}
L_{\cal L} (x) =   x-\langle [{\cal L}] | x \rangle [{\cal L}]  \, ,
\eeq
where the bracket $\langle \dots \rangle$ denotes the intersection number between the two cycles. 
Such transformations satisfy the braid group relations on $n$ strands
\beqa\label{braidrelations}
L_{{\cal L}_i} L_{{\cal L}_j} L_{{\cal L}_i} &=& L_{{\cal L}_j} L_{{\cal L}_i} L_{{\cal L}_j}\,,\quad
{\rm for}\,|i-j|=1\\
L_{{\cal L}_i} L_{{\cal L}_j} &=& L_{{\cal L}_j} L_{{\cal L}_i}\,,\quad{\rm for}\,|i-j|>1\,.\nonumber
\eeqa
Picard Lefschetz transformations play a role in the context of BPS solitons 
(described by A-type D-branes ) in Landau-Ginzburg models \cite{Hori:2000ck}. Here, the intersection numbers are realized as soliton 
numbers, and the Picard Lefschetz monodromy captures their changes under deformations of the superpotential.

A natural question is whether this braid group action extends to the
level of the full topological D-brane category rather than just the charges. This question has been answered in the work of Seidel \cite{Seidel1}, who constructed braid group actions 
on categories of A-branes, which is mathematically described 
by the Fukaya category. Via mirror symmetry, this action should carry over to an action on the mirror
B-brane category. Indeed, this was constructed by Seidel and Thomas in 
\cite{Seidel-Thomas}. The authors introduce the notion of spherical objects $E$ in the derived category of coherent sheaves on the target space manifold $X$ which satisfy
the condition
\beq
\Ext^i(E, E ) = 
 \left\{ \begin{array}{cc} \CC & i=0,n \\ 0 & i\neq 0,n \end{array} \right.\,,
\eeq
where $n$ is the complex dimension of $X$. 
To any such object they associate a Fourier-Mukai transformation which describes
an autoequivalence of the derived category of coherent sheaves on $X$. The kernel of the Fourier-Mukai transformation 
\begin{equation} \label{eq:ConiKernel}
     {\cal K}_{\cal Q}^{\rm C} = 
     {\rm Cone}(r:{\cal Q} \boxtimes {\cal Q}^{\lor} \rightarrow {\cal O}_{\Delta X}) \ ,
\end{equation}
is determined by the large volume complex ${\cal Q}$ associated to the B-brane $Q$ \cite{Aspinwall:2001dz,Seidel-Thomas}. Here ${\cal Q}^\lor$ denotes the dual of ${\cal Q}$. Moreover, ${\cal O}_{\Delta X}$ is the structure sheaf of the diagonal $\Delta X\subset X\times X$, and the
map $r$ is the restriction map to $\Delta X$. If for instance ${\cal Q}=\OO_X$, then the map $r$ restricts $\OO_{X\times X}=\OO_X\boxtimes\OO_X$ to $\OO_{\Delta X}$. Transformations of this type in for example describe the effect on B-branes of monodromies
around conifold points in K\"ahler moduli spaces. These are points where B-branes ${\cal Q}$ become
massless.
The action on the B-brane charges is encoded in the
periods near the conifold point and can be represented by equation (\ref{lefschetz}).

Braid group actions can be obtained provided that there is 
an $A_m$-chain $(E_1,\ldots,E_m)$ of spherical objects. Mimicking  condition (\ref{A_n}) on the A-side, this means 
\beq
\sum \Ext^k(E_i,E_j) = \left\{ \begin{array}{cc} 1 & |i-j|=1 \\ 0 & |i-j| > 1 \end{array} \right.
\eeq

In a different but related context, the braid group has appeared in the context
of 4-dimensional gauge theories with surface operators \cite{Gukov:2006jk,Gukov:2007ck}. 
 
If the non-linear sigma model with target space $X$ has a Landau-Ginzburg phase, the derived category
of coherent sheaves on $X$ is equivalent to the category of B-branes in the corresponding Landau-Ginzburg orbifold, \ie the associated category of equivariant matrix factorizations\footnote{The equivalence can be realized for instance in terms of gauged linear sigma models \cite{Herbst:2008jq}.}. 
In these cases representations of braid groups in the autoequivalences of the derived category of coherent sheaves on $X$ carry over to the respective category of matrix factorizations. 

In the following we will present a world sheet realization of these braid group representations. Namely, we will 
construct defects which satisfy braid relations on the level of their fusion. Since they can be fused with boundary conditions they give rise to functors (in this case autoequivalences) on the respective D-brane categories. One should point out however, that defects have a richer structure than the associated functors on D-brane categories. First of all they are objects in the full conformal field theory, not just in the topologically twisted theories. Moreover, defects can form junctions etc.

The construction we use is rather general and should apply to any $N=(2,2)$-supersymmetric theory. 
It is certainly not limited to theories which have a non-linear sigma model phase. However, we will restrict our considerations to Landau-Ginzburg models, in which everything can be spelled out relatively explicitly.

\subsection{The defects}

The defects that are relevant for us are conifold type defects considered in \cite{Brunner:2008fa}. Following terminology from the work of Seidel and Thomas \cite{Seidel-Thomas} we will call them twist defects. 
Such defects exist in any theory, and in case the theory has $N=(2,2)$ supersymmetry one can lift them to the respective B-twisted model. Namely, in any 
theory we have the trivial identity defect and totally reflective defects. The identity defect $\Id$
is a topological defect and maps via
fusion any boundary condition back to itself. On the other hand, a totally reflective
defect is a defect that provides boundary conditions for each of the two adjacent
theories. In a theory with $N=(2,2)$ supersymmetry we can choose the boundary
conditions to be of B-type, the defect then preserves B-type supersymmetry and can be fused on the level of the B-twisted theory. Choosing for example the boundary condition $P$ on the one side of the defect and its world sheet parity dual $P^*$ on the other side, we
obtain the defect $T_{P,P^*}= P \otimes P^*$. Fusing it with a boundary condition $Q$ one obtains 
a copy of $P$ for every boundary condition changing field between $P$ and $Q$
\beq
T_{P,P^*} * Q = {\cal H}^*(P,Q) \otimes Q\,.
\eeq
Consider now a defect that is a superposition of the identity defect $\Id$ and the totally reflective defect $T_{P,P^*}\left\{ 1\right\}$, \ie $T_{P,P^*}$ shifted by one. There is a universal defect
changing field between $T_{P,P^*}\{1\}$ and $\Id$, which can be used to perturb this configuration.
To see this, note that there is always a bosonic defect changing field between $T_P$ and $\Id$ which
has its origin in the fact that there is an identity field on the boundary condition $P$.
Accordingly, there is a canonical fermion between $T_{P,P^*}\left\{ 1\right\} $ and $\Id$.

This construction mimics the form of the Fourier-Mukai kernel (\ref{eq:ConiKernel}). Here, ${\cal O}_{\Delta X}$ and ${\cal Q}\boxtimes
{\cal Q}^\lor$ correspond to the identity and the purely reflective defects respectively.

Physically, fusion of boundary conditions with this defect mimics how D-branes behave when one 
moves along a closed path in K\"ahler moduli space which encloses a locus $\Delta_P$ on which a D-brane $P$ becomes massless. 
Since copies of $P$ and its anti-brane can be produced at no cost in energy, a probe D-brane $Q$ which is carried around $\Delta_P$ 
forms bound states with $P$ provided there is a suitable tachyon. As we will see explicitly in the next sections this is exactly how fusion with the bound state of $T_{P,P^*}$ and $\Id$ acts on boundary conditions.
The $\Id$ defect preserves a copy of $Q$, whereas the totally reflective 
defect creates as many copies of $P$ as there are tachyons between $P$ and $Q$. Finally the universal
defect changing field between $T_{P,P^*}$ and $\Id$ induces a bound state formation between all the copies of $P$ and the copy of $Q$.

\subsection{Realization in terms of matrix factorizations}

\subsubsection{Twist defects}
The matrix factorizations corresponding to the identity and the totally reflective defects in Landau-Ginzburg models 
have been explicitly described in Section \ref{sec:MF}. 

Since the reflective defect is really a product of two boundary conditions the space 
of defect changing fields between the totally reflective defect $T_{P,P^*}$ and the identity defect $\Id$ 
is isomorphic to the 
space of boundary conditions changing fields between the two boundary conditions
\beq
\HH (T_{P,P^*}, \Id) \equiv \HH (P\otimes P^*, \Id) \equiv \HH(P,P)\,.
\eeq
In particular, via this isomorphism the identity field $\id_P$ on $P$ gives rise to a canonical
defect changing field $\psi_P$
\beq
\HH^0(T_{P,P^*},\Id) \ni \psi_P  \mapsto \id_P \in \HH^0(P,P) \ ,
\eeq
which can be used to perturb the superposition of the (shifted) tensor product and the identity defect. The outcome
\beq
D_P = \cone (\psi_P: T_{P, P^*} \to \Id)
\eeq
of this perturbation obeys some nice universal relations. Indeed, the $D_P$ are the twist defects alluded to above. 
To see this, 
let us first describe  how these defects act on matrix factorizations
\beq
D_P*Q\cong D_P\otimes Q\cong\cone(\psi_P\otimes\id_Q:T_{P,P^*}\otimes Q\rightarrow\Id\otimes Q).
\eeq
As discussed in Section \ref{sec:MF}, $T_{P,P^*}\otimes Q\cong P\otimes\HH(P,Q)$ and $\Id\otimes Q\cong Q$. Indeed, via these isomorphisms, the morphism $\psi_P\otimes\id_Q$ is mapped to the evaluation map $\ev:P\otimes\HH(P,Q)\rightarrow Q$. This can be seen as follows. By definition, the morphism $\psi_P$ is mapped to $\psi_P\mapsto \id_P$ under the isomorphism $\HH(T_{P,P^*},\Id)\cong\HH(P,P)$. Indeed, it also maps to $\psi_P\mapsto \id_{P^*}$ under the isomorphism $\HH(T_{P,P^*},\Id)\cong\HH(P^*,P^*)$, 
and hence $\psi_P\otimes \id_Q\mapsto \id_{P^*}\otimes \id_Q$ under $\HH(T_{P,P^*}\otimes Q,\Id\otimes Q)\cong
\HH(P^*\otimes Q,P^*\otimes Q)$.  Here $P^*\otimes Q$ are matrix factorizations of $0$ and therefore 
ordinary complexes.  But now $\HH(P^*\otimes Q,P^*\otimes Q)\cong\HH(P\otimes P^*\otimes Q, Q)\cong\HH(P\otimes\HH(P,Q),Q)$, and $\id_{P^*}\otimes \id_Q\mapsto \ev$ under this isomorphism, because $\id_{V^*}$ is mapped to the evaluation map under the canonical isomorphism $\Hom(V^*,V^*)\rightarrow(V\otimes V^*)^*$.

Thus, we obtain
\beq\label{twistfusion}
D_P*Q\cong\cone(\ev:P\otimes\HH(P,Q)\rightarrow Q)\,.
\eeq
A similar form can be found for the fusion with the dual of a twist defect which is represented by the dual matrix factorization $D_P^*$. Namely
\beq
D_P^**Q\cong\left(Q^**D_P\right)^*\cong
\left(\cone(\id_{Q^*}\otimes\psi_P:Q^*\otimes T_{P,P^*}\rightarrow Q^*\otimes\Id)\right)^*\,.
\eeq
With arguments similar to the ones above one arrives at 
\beqn
D_P^**Q&\cong&\left(\cone(\ev:\HH(Q,P)\otimes P^*\rightarrow Q^*)\right)^*\nonumber\\
&\cong&\cone(\ev^*:Q\rightarrow(\HH(Q,P))^*\otimes P)\,.\label{dualtwistfusion}
\eeqn
The action \eq{twistfusion} of the defects $D_P$ on matrix factorizations is realized by twist
functors as introduced in \cite{Seidel-Thomas}. There Seidel and Thomas show that under certain assumptions
such twist functors generate representations of braid groups in the groups of autoequivalences of certain derived categories. In the next subsections, we follow their arguments to establish that under similar conditions, twist defects satisfy braid relations with respect to fusion.
\subsubsection{Twist defects for spherical matrix factorizations}
A matrix factorization $P$ is called {\bf n-spherical} if 
\beq\label{conddim}
\HH^i(P,P)=\left\{\begin{array}{ll}\CC\,,& i=0,n\\ 0\,,&{\rm otherwise}\end{array}\right.
\eeq
and for every matrix factorization $Q$, the operator product
\beq\label{non-deg}
\HH^i(P,Q)\otimes\HH^{n-i}(Q,P)\rightarrow\HH^n(P,P)
\eeq
is non-degenerate. Note that we have extended the grading of $\HH^*$ from $\ZZ_2$ to $\ZZ$ by means
of the $R$-charge, and $n$ is chosen such that $-n$ is the $R$-charge anomaly of the disk amplitudes
in the models under consideration\footnote{For models with a realization in terms of non-linear sigma models, $n$ is the complex dimension of the target space.}. Condition \eq{non-deg} is nothing but the non-degeneracy of the boundary two-point function, which holds for unitary theories. We restrict our considerations to such theories and will assume condition \eq{non-deg} in the following.

One nice property of spherical matrix factorizations $P$ is that the associated twist defects $D_P$ are 
indeed group-like \ie
\beq\label{group-like}
D_P*D_P^*=\id=D_P^**D_P\,.
\eeq
In particular, they act on matrix factorization categories as equivalences.
To see this, we calculate the fusion
$D_P*(D_P^**Q)$ for $P$ a spherical and $Q$ any matrix factorization. Using equations 
\eq{dualtwistfusion} and \eq{twistfusion} one arrives at
\beq\label{DD*}
D_P*(D_P^**Q)\cong\cone\left({
\begin{array}{ccc}
P\otimes\HH(P,Q) & \stackrel{f}{\longrightarrow} & \HH(P,P)\otimes(\HH(Q,P))^*\otimes P\\
\mapdown{\ev} & & \mapdown{\ev}\\
Q &\stackrel{\ev^*}{\longrightarrow} & (\HH(Q,P))^*\otimes P
\end{array}}
\right)\,,
\eeq
where the induced map $f$ acts as a dualization $\HH(P,Q)\longrightarrow\HH(P,P)\otimes(\HH(Q,P))^*$
of the operator product $\HH(P,Q)\otimes\HH(Q,P)\rightarrow \HH(P,P)$. Now, since by assumption $P$ is spherical, the non-degeneracy \eq{non-deg} of the operator product $\HH^{n-i}(P,Q)\otimes\HH^i(Q,P)\rightarrow\HH^n(P,P)$ implies that we have an isomorphism
\beq
P\otimes\HH(P,Q)\stackrel{\cong}{\longrightarrow}\HH^n(P,P)\otimes(\HH(Q,P))^*\otimes P
\eeq
in the upper row of \eq{DD*}. Furthermore, in the right column, we have an isomorphism
\beq
\HH^0(P,P)\otimes(\HH(Q,P))^*\otimes P\stackrel{\cong}{\longrightarrow}(\HH(Q,P))^*\otimes P\,.
\eeq
These isomorphisms can be used to reduce the matrix factorization \eq{DD*}  to $Q$. Therefore,
\beq
D_P*(D_P^**Q)\cong Q\,.
\eeq
In a similar way one obtains $D_P^**(D_P*Q)\cong Q$. 

The fact that $D_P$ is group-like for spherical $P$ can be used to show the following relation between twist defects associated to spherical matrix factorizations $P_1$, $P_2$:
\beq\label{sphericalrel}
D_{P_2}*D_{P_1}\cong D_{D_{P_2}P_1}*D_{P_2}\,.
\eeq
For this, we again fuse the equation with a matrix factorization $Q$
\beqn
D_{P_2}*(D_{P_1}*Q)\!\!\!&\cong&\!\!\!\cone\left({
\begin{array}{ccc}
P_2\otimes\HH(P_1,Q)\otimes\HH(P_2,P_1) & \stackrel{\ev}{\longrightarrow} & P_2\otimes\HH(P_2,Q)\\
\mapdown{\ev} & & \mapdown{\ev}\\
P_1\otimes\HH(P_1,Q) &\stackrel{\ev}{\longrightarrow} & Q
\end{array}}
\right)\nonumber\\
&\cong&\!\!\!\cone(\HH(P_1,Q)\otimes D_{P_2}P_1\stackrel{g}{\longrightarrow} D_{P_2}Q)\,.
\eeqn
Since $D_{P_2}$ is group-like $\HH(P_1,Q)\cong\HH(D_{P_2}P_1,D_{P_2}Q)$ and the map
$g$ factors through the evaluation map
\beq
\HH(P_1,Q)\otimes D_{P_2}P_1\stackrel{\cong}{\longrightarrow}\HH(D_{P_2}P_1,D_{P_2}Q)\otimes
D_{P_2}P_1\stackrel{\ev}{\longrightarrow}D_{P_2}Q\,.
\eeq
Thus,
\beqn
D_{P_2}*(D_{P_1}*Q)&\cong&\cone(\HH(D_{P_2}P_1,D_{P_2}Q)\otimes
D_{P_2}P_1\stackrel{\ev}{\longrightarrow}D_{P_2}Q)\nonumber\\
&\cong&D_{D_{P_2}P_1}*(D_{P_2}*Q)\,.
\eeqn
\subsubsection{Defect realization of braid groups}
Relation \eq{sphericalrel} can be used to construct defect realizations of braid groups in the following way. An {\bf $A_m$-sequence of spherical matrix factorizations} is a collection $(P_1,\ldots,P_m)$ of 
spherical matrix factorizations $P_i$ such that
\beq
\dim\HH(P_i,P_j)=\left\{\begin{array}{ll}1\,,& |i-j|=1\\ 0\,& |i-j|>1\end{array}\right.\,.
\eeq
Given such a collection, with the preparations of the last sections, it is easy to see that the associated twist defects $D_{P_i}$ satisfy
braid relations \eq{braidrelations}. Namely for $|i-j|>1$ $\HH(P_i,P_j)=0$, so that from \eq{twistfusion}
one reads off that $D_{P_i}P_j\cong P_j$. Therefore, relation \eq{sphericalrel} implies 
\beq
D_{P_i}*D_{P_j}\cong D_{P_j}*D_{P_i}\quad{\rm for}\; |i-j|>1\,.
\eeq
Moreover, using $\dim\HH(P_{i+1},P_i)=1$ one obtains 
\beqn
D_{P_{i+1}}*P_i&\cong&\cone(P_{i+1}[-r_i]\stackrel{f_{i}\neq 0}{\longrightarrow} P_i)\\
D_{P_{i}}^**P_{i+1}&\cong&\cone(P_{i+1}\stackrel{g_{i}\neq 0}{\longrightarrow} P_i[r_i])
\eeqn
for some shifts $r_i$. But since $\dim\HH(P_{i+1},P_{i})=1$, $f_i$ and $g_i$ are multiples of each other. In particular,
\beq\label{dualrel}
D_{P_2}*P_1[r_i]\cong D_{P_1}^**P_2\,.
\eeq
Now we can conclude the other braid relations:
\beqn
D_{P_i}*D_{P_{i+1}}*D_{P_i}&\cong& D_{P_i}*D_{D_{P_{i+1}}*P_i}*D_{P_{i+1}}\nonumber\\
&\cong&  D_{P_i}*D_{D_{P_i}^**P_{i+1}}*D_{P_{i+1}}\nonumber\\
&\cong& D_{D_{P_i}*D_{P_i}^**P_{i+1}}*D_{P_i}*D_{P_{i+1}}\nonumber\\
&\cong& D_{P_{i+1}}*D_{P_i}*D_{P_{i+1}}\,,.
\eeqn
Here, the first equation is obtained by means of \eq{sphericalrel}.
In the second equation use was made of \eq{dualrel} and the fact that shifts do not change twist
defects. The third equation is again obtained by means of relation \eq{sphericalrel}, where one has to
note that because twist defects of spherical matrix factorizations are group-like, their fusion products with spherical matrix factorizations are still spherical. Finally, in the last equation one again employs \eq{group-like}.

Summarizing, we have established the following. Given a boundary conditions $P$, the superpositions of the shifted purely reflective defects $T_{P,P^*}$ and the trivial defect $\Id$ exhibit a universal defect changing field. The corresponding perturbations lead to twist defects $D_P$, which have some universal properties. For spherical boundary conditions $P$ the associated twist defects $D_P$ are group-like, and with respect to fusion, satisfy the commutation relation \eq{sphericalrel}. Moreover, $A_m$-sequences of boundary conditions give rise to a collection of twist defects, which satisfy braid relations. We have discussed this explicitly in the context of B-type defects in Landau-Ginzburg models, but we expect that the constructions should apply to general $N=(2,2)$-supersymmetric theories. 
\subsection{Examples}
In this section we will present some examples of Landau-Ginzburg models, which exhibit $A_m$-sequences of B-type boundary conditions. By means of the construction above they give rise to 
B-type defects satisfying braid relations.
\subsubsection{Degenerate K3 and fibrations}
Geometrically, the appearance of $A_m$-chains of homology cycles 
has played a prominent role in the discussion of heterotic-IIA duality 
\cite{Witten:1995ex,Aspinwall:1995zi}. In particular, local singularities of K3-surfaces are responsible for the non-abelian 
gauge symmetry enhancement of the IIA string compactified on K3. Via het-IIA duality this is dual to the non-Abelian gauge symmetries appearing at special points in the moduli space of the toroidally compactified weakly coupled heterotic string.
Here, we are particularly interested in the case that the enhanced symmetry is $A_m$, meaning that the K3 should locally
exhibit an orbifold singularity $\CC^2/\ZZ_{m+1}$. Standard examples arise as suitable
hypersurfaces in weighted
projective spaces. Consider for example a degree $12$ hypersurface in $\IP_{(1,3,4,4)}[12]$. Projective equivalence acts on the affine coordinates as
\beq
(x_1, x_2, x_3, x_4) \mapsto (\lambda x_1, \lambda^3 x_2, \lambda^4 x_3, \lambda^4 x_4)
\eeq
For $\lambda^4=1$ this transformation leaves $x_3$ and $x_4$ invariant, leading to a local $\CC^2/\ZZ_4$-singularity in $x_1=0=x_2$. 
This singularity is resolved by three spheres intersecting in an $A_3$ pattern. The hypersurface intersects the singular fixed point set in three points. Hence, there are three $A_3$-chains on this K3 surface. Note also that this means that the Picard lattices of K3's embedded in this weighted projective space generically have 
rank $10$ ($1$ canonical holomorphic $(1,1)$ coming from the hyperplane bundle plus $3\times3$ spheres coming from the resolution of the singularities) so that the embedding requires a restriction to a particular part of the K3 moduli space.

This model has a Landau-Ginzburg orbifold phase with superpotential
\beq\label{hyp}
W=x_1^{12} + x_2^4 + x_3^3 - x_4^3
\eeq
and orbifold group $\Gamma=\ZZ_{12}$. Therefore, one can realize the B-branes supported on the exceptional divisors by $\ZZ_{12}$-equivariant matrix factorizations of $W$. They have been obtained in \cite{Brunner:2006tc}. All the building blocks have already appeared in Sections \ref{sec:mmbranes} and \ref{orbifold}. The relevant matrix factorizations are
\beq
E_m^n = Q^1_{m}(x_1) \otimes Q^1_0(x_2) \otimes P^{\left\{ n \right\}}(x_3,x_4)_0\,.
\eeq
This is a tensor product of three linear matrix factorizations: the ordinary one-variable factorizations $Q^1$ in the first two factors, and a permutation matrix factorization $P^{\{n\}}$ in the last two factors.
Here, $m$ specifies the $\ZZ_{12}$-representation of the matrix factorization, and $n\in\ZZ_3$ denotes
which of the three $A_3$-sequences $E_m^n$ belongs to.
To motivate that these are good candidates for the matrix factorizations realizing the $A_3$-sequences of B-type boundary conditions,
one can use a simplified version of the arguments in \cite{Aspinwall:2006ib,Herbst:2008jq}. 
The latter suggests that 
the B-type boundary conditions associated to $E_m^n$ are localized at the zero locus of the factorization
\beq\label{deg-points}
x_1=x_2=x_3-\eta_n x_4=0 \ ,
\eeq
which is the $\ZZ_4$-singularity. This is blown up by the exceptional divisors, and it is known \eg from orbifold theories, that the fractional branes of the corresponding $\ZZ_4$ represent B-branes wrapping the exceptional divisors at large volume. 

Of coures, one can just check that the $E_{3m}^n$ are spherical and form $A_3$-sequences. The Witten-index between the matrix factorizations $E_{M}^n$ and $E_{N}^n$ is easily calculated to be
\beqn\label{intmat}
I_{M,N}&=&\left((1-g^{-1})(1-g^3)(1+g^4)\right)_{M,N} \\
&=& \left(2-g^{-1}-g^{-3}+g^{-4}+g^4-g^3-g\right)_{M,N},\nonumber
\eeqn
where 
\beq
g_{M,N}=\delta_{M-N,1}^{(12)}
\eeq
is the $\ZZ_{12}$-shift matrix.
Since these matrix factorizations are tensor products of matrix factorizations which do not have bosons and fermions at the same degree, and since furthermore no cancellation occurred in the expansion of 
\eq{intmat} one can indeed even read off the dimensions of
the corresponding open string Hilbert spaces $\HH^*(E^n_{M},E^n_{N})$. One obtains that the $E_m^n$ are spherical and that $(E^n_0,E^n_3,E^n_6)$ constitute $A_3$-sequences of spherical objects. 

Let us turn to models with three dimensional target spaces. Examples that gained particular importance in the context of string
dualities are K3 fibrations. Here, one expects an enhanced gauge symmetry at points in moduli space where the fiber exhibits an ADE degeneration.
For instance, a hypersurface in weighted projective space
$\IP_{1,1,6,8,8}[24]$ is a K3-fibration over $\PP^1$ with fiber a hypersurface in 
$\IP_{(1,3,4,4)}[12]$. To see this, we intersect the hypersurface 
with a linear equation in the coordinates $x_0,x_1$ of weight $1$. 
A special case is $x_0=0$ for which the hypersurface equation reduces to
\beq
x_1^{24}+x_2^4 +x_3^3 -x_4^3=0
\eeq
from which we recover the K3 hypersurface equation \eq{hyp} by substituting $y_2=x_2^2$ which is single valued because of 
quasi-projective equivalence.
In complete analogy to the above discussion, the fibers degenerate at the points (\ref{deg-points}). Hence, one expects an $A_3$-intersection pattern for the matrix factorizations
\beq
\hat{E}_m^n=Q^1_{m}(x_0) \otimes  Q^1_0(x_1) \otimes Q^1_0(x_2) \otimes P^{\left\{ n \right\}}_0(x_3,x_4)\,. 
\eeq
Indeed, the intersection matrix between the $\hat{E}_m^n$ is given by
\beqa\label{intmat2}
I&=&(1-g^{-1})^2(1- g^{-6})(1+ g^8)\\ \nonumber
&=& -2 g^{-1}+ g^{-2}- g^{-6} + 2 g^{-7} -  g^{-8} 
+ g^8-2 g^7+ g^6- g^2+2 g\,,
\eeqa
where now $g$ denotes the $\ZZ_{24}$-shift matrix. Again, no cancellation occured in the expansion of \eq{intmat2}, and one can read off that the matrix factorizations 
$\hat E_m^n$ are spherical, and that $(\hat E_0^n,\hat E_6^n,\hat E_{12}^n)$ constitute $A_3$-sequences.
Many more examples can be constructed in a similar manner, making use of the divisibility patterns of the weights. 

$A_n$-sequences can also be obtained in these examples using tensor products of the one variable matrix factorizations $Q^1$ only.
These factorizations have the advantage that they are universally available in any theory with an $R$-charge, since any quasi-homogeneous superpotential can be factorized as
$W=\sum_i x_i A_i$. Geometrically they come from the embedding quasi-projective space. 

Moreover, one easily sees that
tensor products of linear matrix factorizations $Q^1$ are always spherical: The spectrum for the $i$th factor of the tensor product consists of the identity and one fermion (see \eg Section \ref{sec:mmbranes}). The part of the spectrum of the tensor product factorizations which survives the orbifold projection consists of the identity and the product of the fermionic fields in each factor. Hence condition (\ref{conddim}) is always satisfied. This means in particular that the corresponding twist defects are always group-like and the induced functors on the B-brane categories are 
autoequivalences.
Geometrically, these universal autoequivalences correspond to the monodromy in K\"ahler moduli space around the locus where the highest dimensional D-brane becomes massless.

Making additional assumptions on the divisibility of the weights, one can construct further examples which exhibit $A_n$-sequences of spherical tensor product boundary conditions.

\subsubsection{A non-geometric Landau-Ginzburg example}

A very simple class of Landau-Ginzburg models exhibiting $A_m$-sequences of spherical B-type boundary conditions are the LG-orbifolds with two chiral superfields $x_1$, $x_2$, superpotential
\beq
W= x_1^d + x_2^d 
\eeq
and orbifold group $\Gamma=\ZZ_d$ whose generator acts on the $x_i$ by
\beq\label{orbact}
G:(x_1, x_2) \mapsto (\omega x_1, \omega^{-1} x_2) , \quad \omega = e^{\frac{2\pi i}{d}}
\eeq
The intersection matrix for the linear tensor product factorizations  $F_m=Q^1_m \otimes Q^1_0$ in this model is given by 
\beq
I=(1-g^{-1})(1-g) = 2-g^{-1} -g\,,
\eeq
where $g$ denotes the $\ZZ_d$-shift matrix. As in the previous examples, also here one can read off the 
dimensions of the BRST-cohomologies from $I$.
Any collection of $d-1$ of the $F_m$ forms an $A_{d-1}$-sequence of spherical matrix factorizations.

This model is non geometric in the sense that it has (generically) non-integer central charge and thus no direct geometrical interpretation. Note however its close relationship to the corresponding noncompact models $\CC^2/\ZZ_d$ which can be obtained by setting the superpotential to zero. 

\subsubsection{Hirzebruch-Jung examples}

The examples discussed in the last section can be generalized to the Landau-Ginzburg models
\beq\label{eq:HJLG}
\left(W= x_1^d + x_2^d\right)/ \ZZ_{d(k)}\,.
\eeq
As before, one considers $\ZZ_d$-orbifolds of the Landau-Ginzburg models with two chiral superfiels $x_1$, $x_2$ and superpotential $W=x_1^d+x_2^d$, where now the orbifold generators act as  
\beq
G:(x_1, x_2) \mapsto (\omega x_1, \omega^{k} x_2) , \quad \omega = e^{\frac{2\pi i}{d}}\,.
\eeq
Here, $d$ and $k$ are assumed to be coprime, and the previous examples are obtained by setting $k=d-1$. 

The corresponding non-compact models have appeared in the context of non-super\-symmetric orbifolds $\CC^2/\ZZ_{d(k)}$ \cite{Adams:2001sv,Harvey:2001wm,Martinec:2002wg,Moore:2004yt}. The orbifold singularity can be resolved using the Hirzebruch-Jung resolution, which replaces the singular point by a chain of $\IP_1$'s whose intersection pattern is determined by the continuous fraction expansion
\beq
\frac{d}{k} = a_1 - \frac{1}{a_2 - \frac{1}{a_3- \frac{1}{\dots 1/a_f}}}=
[a_1, \dots , a_f].
\eeq
The $a_i$ are the self-intersection numbers of the $f$ exceptional $\IP_1$'s blown up in the resolution.
The intersection number between subsequent spheres is $1$. Hence, for those $(d,k)$ such that the continuous fraction expansion of $\frac{d}{k}$ contains a string $a_s=a_{s+1}=\ldots=a_{s+r-1}=2$ 
the corresponding orbifold model contains an $A_r$-sequence of spherical B-branes wrapping the associated exceptional $\PP_1$'s.

Because of the close relationship between non-compact orbifold and Landau-Ginzburg models
such $A_r$-chains must also be present in the LG models (\ref{eq:HJLG}). As an example, let us consider the case that the string of $2$'s is located at the beginning of the continuous fraction expansion of $\frac{d}{k}$, followed by one further integer $b>2$:
\beq\label{eq:HJexample}
\frac{d}{k} = [2,\dots,2, b]= \frac{(r+1)b-r}{rb-(r-1)}\,.
\eeq
As candidates for the $A_r$-sequence of spherical matrix factorizations we again choose tensor products
\beq\label{eq:HJbranes}
G_m= Q_m^\Delta \otimes Q_0^1, \quad \Delta=d-k\,.
\eeq
Quite generally, if the continued fraction expansion of $\frac{d}{k}$ starts with a $2$, we can conclude that $\Delta \leq \frac{d}{2}$. 
In the specific case (\ref{eq:HJexample}) $\Delta=b-1$. 
The intersection matrix of the $G_m$ is given by 
\beqa\label{eq:HJintersect}
I&=& \left( \sum_{j=-\Delta}^{\Delta -1} sgn(j) g^j \right)\left(1-g^{-k}\right) \nonumber\\ \label{specialIintersect}
&=& 2\sum_{j=0}^{\Delta -1} g^j- \sum_{j=\Delta}^{2\Delta -1}g^j - \sum_{j=-\Delta}^{-1} g^j \ ,
\eeqa
where 
$g$ is the $d$-dimensional shift matrix. Note that no cancellation occurs in the expansion \eq{eq:HJintersect}. Since the $G_m$ are tensor products of matrix factorizations which do not have bosons and fermions at the same degree, one can therefore read off the dimensions of the BRST cohomology groups directly from $I$. One finds that the $G_m$ are spherical and that $\dim\HH^*(G_{a\Delta},G_{b\Delta})=1$ for $|a-b|=1$. Indeed, the $G_{a\Delta}$ with $a\in\{0,\ldots, r-1\}$ constitute an $A_r$-sequence of spherical matrix factorizations. Namely, since $(r-1)\Delta<d$, no summand $g^{n\Delta}$ with $r>|n|\geq 2$ appears in \eq{eq:HJintersect}, and hence $\dim\HH^*(G_{a\Delta},G_{b\Delta})=0$ for all $a,b\in\{0,\ldots, r-1\}$ with $|a-b|>1$. 

In exactly the same way one obtains $A_r$-sequences of spherical matrix factorizations for models in which the 
single integer $b$ in the continuous fraction expansion  \eq{eq:HJexample} is replaced by an arbitrary string
\beq
\frac{d}{k} = [2,\dots, 2, b_{r+1}, \dots, b_f] \ .
\eeq
The discussion of the general case 
\beq
\frac{d}{k}=[a_1, \dots, a_i, 2,\dots, 2,b_{r+i+1}, \dots , b_f]
\eeq
is slightly more involved and can be found in \cite{rossi}.

\subsection*{Acknowledgements}
D.~R.~ is supported by a DFG research fellowship and partially by DOE-grant DE-FG02-96ER40949.
I.~B. is supported by a EURYI award of the European Science Foundation.
We would like to thank Matthias Gaberdiel and Ingo Runkel for stimulating discussions.

\bibliographystyle{flowdef}
\bibliography{flowdef}

\end{document}